%% file: root.tex
\documentclass[letterpaper, 10 pt, conference]{ieeeconf}  

\IEEEoverridecommandlockouts                              

\usepackage{amsmath}
\usepackage{nicefrac}
\usepackage{units}
\usepackage[pdftex]{graphicx}
\usepackage{import}
\usepackage{xcolor}
\usepackage{colortbl}
\usepackage{caption}
\usepackage{multirow}
\usepackage{cases}
\usepackage{booktabs}
\title{\LARGE \bf
   Human-Based Risk Model for Improved Driver Support \\ in Interactive Driving Scenarios  
}

\author{Tim Puphal$^{1,2}$, Benedict Flade$^1$, Matti Kr\"uger$^1$, Ryohei Hirano$^3$ and Akihito Kimata$^3$ 
\thanks{$^1$ Honda Research Institute Europe, Carl-Legien-Str. 30, 63073 Offenbach, Germany. Email: {\tt\footnotesize \{firstname.lastname\}@honda-ri.de}}
\thanks{$^2$ Honda Research Institute Japan, 8-1 Honcho, Wako, 351-0114 Saitamam Japan. Email: {\tt\footnotesize \{tim.puphal\}@jp.honda-ri.com}}
\thanks{
$^3$ Honda R$\&$D Co., Ltd. 4630 Shimotakanezawa, 321-3393 Tochigi, Japan. Email: {\tt\footnotesize \{firstname$\_$lastname\}@jp.honda}}
}

\begin{document}

\maketitle
\thispagestyle{empty}
\pagestyle{empty}

\begin{abstract}
This paper addresses the problem of human-based driver support. Nowadays, driver support systems help users to operate safely in many driving situations. Nevertheless, these systems do not fully use the rich information that is available from sensing the human driver. In this paper, we therefore present a human-based risk model that uses driver information for improved driver support. In contrast to state of the art, our proposed risk model combines a) the current driver perception based on driver errors, such as the driver overlooking another vehicle (i.e., notice error), and b) driver personalization, such as the driver being defensive or confident. In extensive simulations of multiple interactive driving scenarios, we show that our novel human-based risk model achieves earlier warning times and reduced warning errors compared to a baseline risk model not using human driver information. 
\end{abstract}

\input{chapters/intro.tex}

\input{chapters/risk_model.tex}
\input{chapters/human_based_risk_model.tex}

\input{chapters/experiments.tex}

\input{chapters/conclusion.tex}

\addtolength{\textheight}{-12cm}   
                

\bibliographystyle{IEEEtran}
\bibliography{bib}

\end{document}

%% file: chapters/intro.tex
\vspace{-0.05cm}
\section{Introduction}
Driver support systems are nowadays successfully helping users in many everyday driving situations \cite{bengler2014}. For example, adaptive cruise control \cite{yinglong2019} has been introduced in production vehicles, using the time headway to the front driving car as a risk model to successfully follow other vehicles. Various parking assistance systems \cite{heimberger2017} are standard in vehicles, using distance signals to inform the driver about nearby objects like buildings and walls. Finally, collision mitigation systems, such as the work of \cite{bengtsson2010}, are also widespread in the market, using time-to-collision metrics to warn about vehicles in the forward driving area. 
 
However, until now, these driver support systems mostly apply only risk models and do not consider the human driver. According to previous research \cite{ortega2020}, rich information is available from sensing the human driver. There are already first approaches, see, for instance, the work \cite{fridman2018}, which check if the driver is falling asleep or which ensure that the driver is not looking at their phone while driving to improve the system usability. Using such human driver information has a large potential to improve driver support.

\begin{figure}[t!]
  \centering
  \vspace*{0.3cm}
  \resizebox{0.94\linewidth}{!}{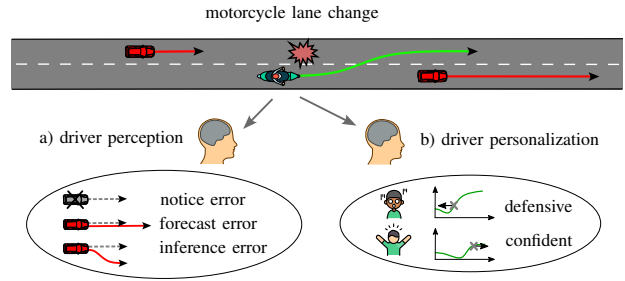}
  \vspace*{0.09cm}
  \caption[]{The image shows a driving situation example of a dynamic lane change in which human factors can change the preferred diver support. In this paper, we present a human-based risk model that combines a) the driver perception based on driver errors (i.e., notice error, forecast error and inference error) and b) driver personalization based on driver types (e.g., defensive and confident).}
  \label{fig:intro}
  \vspace{-0.01cm}
\end{figure}

Fig. \ref{fig:intro} shows an example of a driving situation in which human driver information can improve driver support. In this example, a motorcycle rider intends to change lanes because of a slower car in front. The rider must consider the car on the neighboring lane to successfully change lanes. Here, the driver perception influences the future driving risk. We define \textit{driver perception} based on driver errors. For example, if the motorcycle rider overlooks the car on the neighboring lane (i.e., notice error), the rider might make an unexpected and dangerous lane change. Other driver errors include the rider estimating a wrong velocity of the other car (i.e., forecast error) or inferring a wrong lane change intention of the other car (i.e., inference error). Driver perception modeling allows warnings tailored to the driver's current state.

Additionally, the motorcycle rider can exhibit different driver types (e.g., defensive and confident) in the given driving situation. Such human driver information changes the warning preference for driver support. A defensive motorcycle rider, for example, prefers an early warning to have enough time reacting to the situation. In contrast, a confident rider prefers a late warning. \textit{Driver personalization} based on driver types can also improve driver support.

In this paper, we present a human-based risk model that improves driver support by combining both aformentioned human driver aspects: a)~the current driver perception and b)~driver personalization. Our proposed approach is based on a stochastic collision risk model.
In extensive simulations of multiple interactive driving scenarios, such as the depicted lane change situation, we show that our approach achieves improved driver support. On average, the human-based risk model achieves an earlier warning time and reduced warning errors compared to a state-of-the-art baseline risk model that does not use human driver information.

\subsection{Related Work}

In recent human factors research, there are many technologies proposed that are able to sense the human driver. For example, the driver's gaze, sleepiness and general attention for the driving situation can be estimated from cameras inside the vehicle pointed towards the driver \cite{fridman2018}. In addition, motion data from the driver's vehicle has been shown to correlate with the general driving type of the driver \cite{ingelder2016}. Still, only few approaches use human driver information to improve driver support.

Modeling driver perception is here challenging and has not received much attention in the context of driver support. One interesting work in this direction is \cite{mccall2007}, in which a system is proposed that detects if the driver intends to brake and, if so, issues no warning of the upcoming critical objects. Another related work is \cite{wang2020}, which took another approach by allowing the driver to see the current system predictions. The driver can then correct the prediction for automated driving through speech input. 

In the context of personalization for driver support, there are many approaches presented that estimate parameters of vehicle control systems or driver models, but there are only few approaches that use personalization to improve driver support in the form of risk warnings. As an example, the authors of \cite{orth2017} estimate driver-specific critical intersection merging gaps from driving data with machine learning. The authors of \cite{kreutz2022} estimate reaction times for the driver model called Intelligent Driver Model (IDM), and the authors of the paper \cite{kolekar2020} adapt parameters for the driver's risk field to different driving situations, such as curve driving, and could safely handle the scenarios when coupling the system to a vehicle controller. 

Lastly, a driver perception model was previously developed with risk considerations for improved driver warning, as we have demonstrated in \cite{puphal2023}. Furthermore, we showed that driver support with personalization can effectively adapt risk warning to the driver in \cite{puphal2024}. In this paper, we combine both aspects of human driver information for driver support: a)~the current driver perception and b) driver personalization. To the best of the authors' knowledge, this is the first time such a combined human-based risk model is proposed.

\subsection{Contribution} 
In this paper, we build upon related work and present a human-based risk model to achieve improved driver support. In detail, the contributions are two-fold: 
\begin{enumerate}
 \item We show how human models of driver perception and driver personalization can be combined into a human-based risk model for driver support. 
 \item We present a large-scale simulation experiment and analysis for multiple driving scenarios with variations of the scenarios and of the human driver information input for the human-based risk model. 
\end{enumerate}
In the experiments, we assume sensed human driver information as given and focus on how the information can be used for risk models. The results highlight that human factors can improve driver support for many driving scenarios.

The remainder of the paper is structured as follows. In the next Section \ref{sec:risk_model}, we first present state-of-the-art risk models and driver behavior planning. Section \ref{sec:risk_model} then introduces the human-based risk model and explains the system architecture and parametrization. Section \ref{sec:experiments} shows evenutally the experimental setup and the statistical results of applying the proposed human-based risk model in simulation and Section~\ref{sec:conclusion} gives a conclusion and outlook.

%% file: images/1_driving_scenario_example2.pdf_tex
\begingroup%
  \makeatletter%
  \providecommand\color[2][]{%
    \errmessage{(Inkscape) Color is used for the text in Inkscape, but the package 'color.sty' is not loaded}%
    \renewcommand\color[2][]{}%
  }%
  \providecommand\transparent[1]{%
    \errmessage{(Inkscape) Transparency is used (non-zero) for the text in Inkscape, but the package 'transparent.sty' is not loaded}%
    \renewcommand\transparent[1]{}%
  }%
  \providecommand\rotatebox[2]{#2}%
  \newcommand*\fsize{\dimexpr\f@size pt\relax}%
  \newcommand*\lineheight[1]{\fontsize{\fsize}{#1\fsize}\selectfont}%
  \ifx\svgwidth\undefined%
    \setlength{\unitlength}{340.21773499bp}%
    \ifx\svgscale\undefined%
      \relax%
    \else%
      \setlength{\unitlength}{\unitlength * \real{\svgscale}}%
    \fi%
  \else%
    \setlength{\unitlength}{\svgwidth}%
  \fi%
  \global\let\svgwidth\undefined%
  \global\let\svgscale\undefined%
  \makeatother%
  \begin{picture}(1,0.43948971)%
    \lineheight{1}%
    \setlength\tabcolsep{0pt}%
    \put(0.31697499,0.42754461){\color[rgb]{0,0,0}\makebox(0,0)[lt]{\lineheight{1.25}\smash{\begin{tabular}[t]{l}motorcycle lane change\end{tabular}}}}%
    \put(0.24347929,0.1209908){\color[rgb]{0,0,0}\makebox(0,0)[lt]{\lineheight{1.25}\smash{\begin{tabular}[t]{l}notice error\end{tabular}}}}%
    \put(0.24182317,0.0821781){\color[rgb]{0,0,0}\makebox(0,0)[lt]{\lineheight{1.25}\smash{\begin{tabular}[t]{l}forecast error\end{tabular}}}}%
    \put(0.24171482,0.04162313){\color[rgb]{0,0,0}\makebox(0,0)[lt]{\lineheight{1.25}\smash{\begin{tabular}[t]{l}inference error\end{tabular}}}}%
    \put(0.80695184,0.10912353){\color[rgb]{0,0,0}\makebox(0,0)[lt]{\lineheight{1.25}\smash{\begin{tabular}[t]{l}defensive\end{tabular}}}}%
    \put(0.80872332,0.05434308){\color[rgb]{0,0,0}\makebox(0,0)[lt]{\lineheight{1.25}\smash{\begin{tabular}[t]{l}confident\end{tabular}}}}%
    \put(0.66885711,0.21932534){\color[rgb]{0,0,0}\makebox(0,0)[lt]{\lineheight{1.25}\smash{\begin{tabular}[t]{l}b) driver personalization\\\end{tabular}}}}%
    \put(0.04516779,0.25844816){\color[rgb]{0,0,0}\makebox(0,0)[lt]{\lineheight{1.25}\smash{\begin{tabular}[t]{l}\\a) driver perception\\\end{tabular}}}}%
    \put(0,0){\includegraphics[width=\unitlength,page=1]{1_driving_scenario_example2.pdf}}%
  \end{picture}%
\endgroup%

%% file: chapters/risk_model.tex
\section{Risk Model}
\label{sec:risk_model}
\begin{figure}[t!]
  \centering
  \vspace*{0.03cm}
  \resizebox{0.96\linewidth}{!}{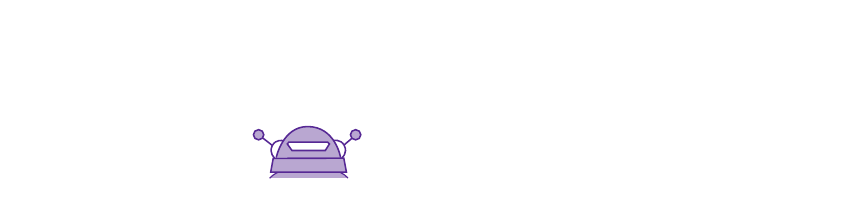}
  \vspace*{-0.17cm}
  \caption[]{Common risk models do not consider the human driver for driver support. The risk model uses only the vehicle states in the driving situation and generates a warning signal.}
  \label{fig:baseline_model}
\end{figure}

Common risk models in state of the art do not consider the human driver but use only the vehicle states to generate a warning. Fig. \ref{fig:baseline_model} shows the block diagram of such a risk model. The vehicle states include the vehicle state of the driver and the state of surrounding vehicles. 
The model then predicts driving risks, e.g., vehicle collisions, by assuming a constant velocity for both the driver and other vehicles and predicting their positions along driving paths. The resulting risk value is used to generate a warning signal. 

Since predicting into the future involves uncertainty, many common risk models account additionally for uncertainties in the vehicle trajectories. In this paper, we use our previously published approach \cite{puphal2019} for this purpose. 

The risk model includes the stochastic models of the Gaussian method and survival analysis. The Gaussian method includes uncertainty in vehicle positions by modeling two-dimensional Gaussian distributions. The distributions are defined by mean positions $\boldsymbol{\mu}_{1}$ and $\boldsymbol{\mu}_{2}$, and uncertainty matrices $\mathbf{\Sigma}_{1}$ and $\mathbf{\Sigma}_{2}$~for the driver's vehicle and another vehicle. In this way, the collision probability between the two vehicles can be given by the overlap of the distributions. We write for the collision probability 
 \begin{align}
P_{\text{coll}}(t+s) = \operatorname{det}|2\pi&(\mathbf{\Sigma}_{1}+\mathbf{\Sigma}_{2})|^{-\frac{1}{2}} * \nonumber \\ \exp\{-\frac{1}{2}
(\boldsymbol{\mu}_{2}-\boldsymbol{\mu}_{1})^T(\mathbf{\Sigma}_{1}&+\mathbf{\Sigma}_{2})^{-1}
(\boldsymbol{\mu}_{2}-\boldsymbol{\mu}_{1})\},
\label{eq:prodgauss}
\end{align}
which is computed for each future timestep $t+s$ starting from the current timestep $t$ in the driving prediction. 

In the Gaussian method, the two-dimensional Gaussian distributions are defined to grow over the future time. The further the risk model predicts into the future, the less likely the position estimates of the vehicles become.

Finally, the risk model uses the survival analysis to condense the prediction into one risk value for the current time. In detail, we integrate the collision probabilities $P_{\text{coll}}(t+s)$ over the future time. The final risk value is given by
\begin{equation}\label{IntRisk}
R(t) \sim \int_0^{s_{\text{max}}} P_{\text{coll}}(t+s)S(s;t)\,ds,
\end{equation} 
in which we cap the integration of the survival analysis with a fixed prediction horizon $s_{\text{max}}$. Moreover, a survival function $S(s;t)$ is added in the integration to include uncertainty in time. The function resembles a decaying exponential function and rates collisions in the distant future less than those in the near future of the risk prediction. 

\subsection{Risk Maps}
\label{subsec:riskmaps}
\begin{figure}[t!]
  \centering
  \vspace*{-0.13cm}
  \resizebox{0.96\linewidth}{!}{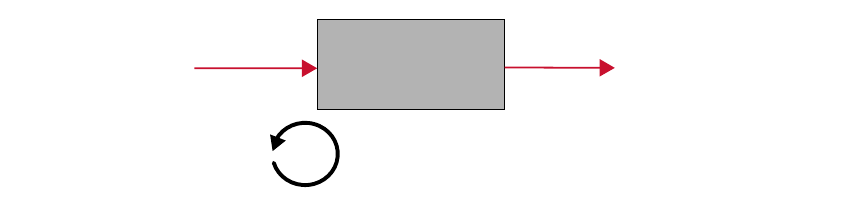}
  \vspace*{-0.2cm}
  \caption[]{Risk Maps for behavior planning. The risk model can be used in a cost function to find an optimal behavior.}
  \label{fig:risk_maps}
\end{figure}

While the risk model allows to generate a warning signal for driver support, the risk model can also be used in driver behavior planning. We will briefly describe state-of-the-art behavior planning using risk models. Fig. \ref{fig:risk_maps} shows the block diagram of the example approach Risk Maps \cite{puphal2022}.

To plan a behavior, Risk Maps initially samples different target velocities for the vehicle of the driver, representing acceleration and deceleration behaviors along driving paths. The target velocity with minimal driving costs is then the optimal driving behavior. The considered costs consist here of the collision risk $R$, as well as the driver utility $U$ (i.e., time spent to reach goal) and driver comfort $O$ (i.e., number of velocity changes). In this way, the planning problem can be formulated as 
\begin{align}
C(v^h) = R(v^h) - U(v^h) &+ O(v^h), \\ \text{with } 
v_{\text{tar}} = \text{argmin}_{h} C(v^h).
\end{align}
The output is an optimal behavior in the form of the target velocity $v_{\text{tar}}$ selected from sampled velocities $v^h$, and the risk model is used for each sampled motion behavior.

%% file: images/2_baseline_risk_model.pdf_tex
\begingroup%
  \makeatletter%
  \providecommand\color[2][]{%
    \errmessage{(Inkscape) Color is used for the text in Inkscape, but the package 'color.sty' is not loaded}%
    \renewcommand\color[2][]{}%
  }%
  \providecommand\transparent[1]{%
    \errmessage{(Inkscape) Transparency is used (non-zero) for the text in Inkscape, but the package 'transparent.sty' is not loaded}%
    \renewcommand\transparent[1]{}%
  }%
  \providecommand\rotatebox[2]{#2}%
  \newcommand*\fsize{\dimexpr\f@size pt\relax}%
  \newcommand*\lineheight[1]{\fontsize{\fsize}{#1\fsize}\selectfont}%
  \ifx\svgwidth\undefined%
    \setlength{\unitlength}{310.17073059bp}%
    \ifx\svgscale\undefined%
      \relax%
    \else%
      \setlength{\unitlength}{\unitlength * \real{\svgscale}}%
    \fi%
  \else%
    \setlength{\unitlength}{\svgwidth}%
  \fi%
  \global\let\svgwidth\undefined%
  \global\let\svgscale\undefined%
  \makeatother%
  \begin{picture}(1,0.24369852)%
    \lineheight{1}%
    \setlength\tabcolsep{0pt}%
    \put(0,0){\includegraphics[width=\unitlength,page=1]{2_baseline_risk_model.pdf}}%
    \put(0.44114384,0.07208238){\color[rgb]{0,0,0}\makebox(0,0)[lt]{\lineheight{1.25}\smash{\begin{tabular}[t]{l}-predicts future driving risks \\-e.g., vehicle collisions\end{tabular}}}}%
    \put(0,0){\includegraphics[width=\unitlength,page=2]{2_baseline_risk_model.pdf}}%
    \put(0.61377701,0.19965299){\color[rgb]{0,0,0}\makebox(0,0)[lt]{\lineheight{1.25}\smash{\begin{tabular}[t]{l}warning signal\end{tabular}}}}%
    \put(0.39617105,0.15348999){\color[rgb]{1,1,1}\makebox(0,0)[lt]{\lineheight{1.25}\smash{\begin{tabular}[t]{l}\textbf{Risk model}\end{tabular}}}}%
    \put(0,0){\includegraphics[width=\unitlength,page=3]{2_baseline_risk_model.pdf}}%
    \put(0.16140134,0.19691805){\color[rgb]{0,0,0}\makebox(0,0)[lt]{\lineheight{1.25}\smash{\begin{tabular}[t]{l}vehicle states\end{tabular}}}}%
  \end{picture}%
\endgroup%

%% file: images/2_baseline_Risk_Maps.pdf_tex
\begingroup%
  \makeatletter%
  \providecommand\color[2][]{%
    \errmessage{(Inkscape) Color is used for the text in Inkscape, but the package 'color.sty' is not loaded}%
    \renewcommand\color[2][]{}%
  }%
  \providecommand\transparent[1]{%
    \errmessage{(Inkscape) Transparency is used (non-zero) for the text in Inkscape, but the package 'transparent.sty' is not loaded}%
    \renewcommand\transparent[1]{}%
  }%
  \providecommand\rotatebox[2]{#2}%
  \newcommand*\fsize{\dimexpr\f@size pt\relax}%
  \newcommand*\lineheight[1]{\fontsize{\fsize}{#1\fsize}\selectfont}%
  \ifx\svgwidth\undefined%
    \setlength{\unitlength}{310.17073059bp}%
    \ifx\svgscale\undefined%
      \relax%
    \else%
      \setlength{\unitlength}{\unitlength * \real{\svgscale}}%
    \fi%
  \else%
    \setlength{\unitlength}{\svgwidth}%
  \fi%
  \global\let\svgwidth\undefined%
  \global\let\svgscale\undefined%
  \makeatother%
  \begin{picture}(1,0.24369852)%
    \lineheight{1}%
    \setlength\tabcolsep{0pt}%
    \put(0,0){\includegraphics[width=\unitlength,page=1]{2_baseline_Risk_Maps.pdf}}%
    \put(0.39581477,0.18269339){\color[rgb]{1,1,1}\makebox(0,0)[lt]{\lineheight{1.25}\smash{\begin{tabular}[t]{l}\textbf{Risk Maps}\\\textbf{planning}\end{tabular}}}}%
    \put(0.4282261,0.07252072){\color[rgb]{0,0,0}\makebox(0,0)[lt]{\lineheight{1.25}\smash{\begin{tabular}[t]{l}-samples motion behaviors\\-uses risk model \end{tabular}}}}%
    \put(0.61377701,0.19965299){\color[rgb]{0,0,0}\makebox(0,0)[lt]{\lineheight{1.25}\smash{\begin{tabular}[t]{l}optimal behavior\end{tabular}}}}%
    \put(0.16140134,0.19691805){\color[rgb]{0,0,0}\makebox(0,0)[lt]{\lineheight{1.25}\smash{\begin{tabular}[t]{l}vehicle states\end{tabular}}}}%
  \end{picture}%
\endgroup%

%% file: chapters/human_based_risk_model.tex
\section{Human-Based Risk Model}
\label{sec:human_based_risk_model}
The last section introduced risk models and behavior planning in state of the art. Both approaches do not consider the human driver. In this section, we therefore present a human-based risk model that improves driver support by explicitly including human driver information. 

Fig. \ref{fig:human_based_model} depicts the block diagram of the human-based risk model of this paper. In addition to the risk model, the human-based risk model includes two further components: perceived Risk Maps and risk factor personalization. Perceived Risk Maps (grey block) is a driver behavior planner like the previously mentioned Risk Maps planner but considers vehicle states and driver errors in the planning to model the driver perception. The risk model (blue block) uses the planned driver behavior of perceived Risk Maps. Concretely, the risk is computed by predicting the vehicle of the driver with the found driver behavior of perceived Risk Maps and the other vehicles with a general constant velocity assumption. 

The risk factor personalization (grey block) allows finally to further adapt the driver support to the current driver type. Perceived Risk Maps and the risk model is additionally parametrized based on the risk factor personalization component. 

In this paper, we focus on using sensed human factors for human driver modeling 
and improved driver warning support. We will therefore assume a driver state estimation module that senses human
factors as given. For approaches of state estimation modules, please refer, for example, to the work of \cite{fridman2018}. Estimating the driver state from sensors is another research work by itself.

\subsection{Perceived Risk Maps}
\label{sec:alpha_estimation}

Perceived Risk Maps uses the three driver errors of notice error, forecast error and inference error that describe the current perception of the driver. A notice error models the condition that the driver is unaware of another object. If the driver has not seen the other object, or has seen it more than a fixed time ago, the driver might not be aware of this object. We define the interface of the notice error for perceived Risk Maps using
\begin{equation}
o_{\text{per}}=
    \begin{cases}
        \text{aware,} \hspace{1.074cm} \text{for } \text{NE} \in [0, 0.5), \\
        \text{not aware,} \hspace{0.5cm} \text{for } \text{NE} \in [0.5, 1], \\
    \end{cases}
\label{eq:DE}
\end{equation}
with the discrete variable $o_{\text{per}}$ describing the perceived awareness for the object and an estimated value of the notice error \text{NE} from a state estimation module. The driver can be either aware or not aware of the other object. These variables and all following variables are dynamic and defined for the current time $t$, which gives us $o_{\text{per}}(t)$ and $\text{NE}(t)$. 

\begin{figure}[t!]
  \vspace*{0.22cm}
  \centering
  \resizebox{0.96\linewidth}{!}{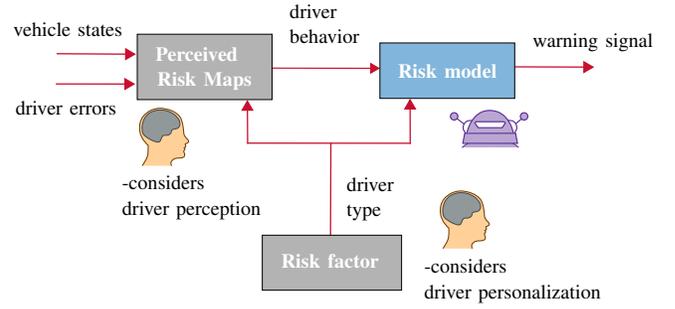}
  \vspace{0.04cm}
  \caption[]{The proposed human-based risk model. Perceived Risk Maps includes the driver perception using driver errors and a risk factor is used for driver personalization. This allows the risk model to include the human driver for driver support.}
  \vspace{-0.1cm}
  \label{fig:human_based_model}
\end{figure}

A forecast error represents the condition that the driver estimates the velocity of another vehicle incorrectly. We thus define the interface of the forecast error as 
\begin{equation}
v_\text{per} = v_{\text{obj}} + \text{FE} \cdot v_{\text{off}}, \hspace{0.5cm}\text{for FE} \in [0, 1],
\end{equation}
in which the contained $v_{\text{per}}$ is the perceived velocity of another vehicle for the driver and $\text{FE}$ represents the forecast error from a state estimation module. The error defines how much an offset value $v_{\text{off}}$ changes the correct, objective velocity $v_{\text{obj}}$, whereby $v_{\text{off}}$ is a parameter and can be chosen depending on the driver state.  

Lastly, an inference error describes a driver that predicts the intention of another vehicle (i.e., driving path) wrong. We define the interface of the inference error as 
\begin{equation}
\mathbf{p}_\text{per} =     
\begin{cases}
    \mathbf{p}_{\text{obj}}, \hspace{0.616cm} \text{for IE} \in [0, 0.5), \\
    \mathbf{p}_{\text{pred}}, \hspace{0.5cm} \text{for IE} \in [0.5, 1]. \\
\end{cases}
\end{equation}
In the equation, two paths are here given. First, the correct, objective future path of the other vehicle $\mathbf{p}_{\text{obj}}$ and second, the predicted path from the ego driver for the other vehicle $\mathbf{p}_{\text{pred}}$. Depending on the estimated value of the inference error $\text{IE}$ from a driver state estimation module, the perceived path from the driver $\mathbf{p}_{\text{per}}$ might be wrong. 

In case of only one path for the other vehicle, we model in the inference error that the driver can also predict a wrong acceleration behavior for the other vehicle with additional parameters of $a_{\text{intent}}$ and $t_{\text{intent}}$ describing the acceleration value and duration of the behavior, respectively. 

The aforementioned driver errors are an input for perceived Risk Maps. The behavior planner perceived Risk Maps uses the vehicle states and the driver errors to plan a driver behavior. The planning equation of Risk Maps from Section \ref{subsec:riskmaps} therefore changes to 
\begin{align}
v_{\text{tar}} = \text{argmin}_{h} C(v^h; o_{\text{per}}, v_\text{per}, \mathbf{p}_\text{per})
\end{align}
with the driver perception formulated according to the interface variables $o_{\text{per}}$, $v_\text{per}$ and $\mathbf{p}_\text{per}$ of the driver errors.

Fig. \ref{fig:human_based_model} shows the inputs and outputs for the component of perceived Risk Maps within the human-based risk model. Using the driver errors for the behavior planner, perceived Risk Maps allows to consider the driver perception. 

\vspace{0.11cm}
\subsection{Risk Factor}
The risk factor defines the driver types for driver personalization. Besides the driver types of a defensive driver and confident driver from the introduction of this paper, we also consider a normal driver type in this paper that lies in between the two driver types.

In the risk model, the risk factor is defined as an additional parameter for the Gaussian method. We scale the maximum assumed uncertainties $\mathbf{\Sigma}_{1}$ and $\mathbf{\Sigma}_{2}$ of the two-dimensional Gaussian distributions in the vehicle predictions with the risk factor~$\alpha$. For example, a defensive driver holds a high value for the risk factor. A defensive driver sees high uncertainty in the future behavior of vehicles and prefers to keep distance to other vehicles. In the same manner, a normal driver holds a medium-sized value and a confident driver holds a low value for the risk factor. While a normal driver keeps reasonable distances to other vehicles, a confident driver keeps low distances to other vehicles, both based on the amount of assumed uncertainty from the risk factor. 

We define the collision probability $P_{\text{coll}}(t, \alpha)$ after personalization as
\begin{align}
P_{\text{coll}}(t+s; \alpha) \sim \operatorname{det}|2\pi&(\mathbf{\Sigma}_{1}(\alpha)+\mathbf{\Sigma}_{2}(\alpha))|^{-\frac{1}{2}} * \nonumber \\ \exp\{-\frac{1}{2}
(\boldsymbol{\mu}_{2}-\boldsymbol{\mu}_{1})^T(\mathbf{\Sigma}_{1}(\alpha)&+\mathbf{\Sigma}_{2}(\alpha))^{-1}
(\boldsymbol{\mu}_{2}-\boldsymbol{\mu}_{1})\}.   
\label{eq:prodgauss}
\end{align}
The equation denotes the uncertainty dependencies to the risk factor~$\alpha$ with $\mathbf{\Sigma}_1(\alpha)$ and $\mathbf{\Sigma}_2(\alpha)$. For better readability, we did not write out the dependency of the mean positions and uncertainties on the prediction time $t+s$.

In a last step, the personalization component considers a further risk weight that is computed based on the risk factor. The risk weight $w_{\alpha}$ adapts the strength of the warning signal to match the driver's expectation. The weight is linearly dependent on the risk factor and leads for a defensive driver to overestimate the risk ($w_{\alpha}>1$) and for a confident driver to underestimate the risk ($w_{\alpha}>1$) compared to a normal driver ($w_{\alpha}~=~1$). We write accordingly
\begin{equation}
W(t) = w_{\alpha} * R(t, \alpha, v_\text{tar}). 
\label{eq:warning_signal}
\end{equation}
for the warning signal. The warning signal $W(t)$ is overall dependent on the risk weight $w_{\alpha}$, the risk factor $\alpha$ and the found driver behavior $v_{\text{tar}}$ from perceived Risk Maps.

Fig. \ref{fig:human_based_model} shows the driver type as an output of the risk factor component. The risk factor personalization allows to include driver personalization in the human-based risk model and further improves the driver support.

\vspace{0.17cm}
\subsection{Parametrization}

The human-based risk model has several parameters that can be chosen by the designer of the driver support system. We will lastly give an overview of these parameters and show example parameter values.
A summary of the parameters is shown in Tab. \ref{tab:parameters}. 

In perceived Risk Maps, three parameters are given that describe the driver errors. For the forecast error, a vehicle velocity offset $v_{\text{off}}$ was introduced. We set the value in this paper to $v_{\text{off}}\hspace{-0.03cm}=\hspace{-0.03cm}-\unit[4]{m/s}$ as an example value for inner-city driving. For the inference error, the parameters of a wrongly assumed acceleration behavior for another vehicle were used with $a_{\text{intent}}$ and $t_{\text{intent}}$. In this paper, we set these parameters to $a_{\text{intent}}\hspace{-0.03cm}=\hspace{-0.03cm}-\unit[1.5]{m/s^2}$ and $t_{\text{intent}}\hspace{-0.03cm}=\hspace{-0.03cm}\unit[3]{s}$. The parameter values for perceived Risk Maps can potentially also be estimated with a driver state estimation module during real driving. 

Regarding the driver types, the personalized risk factor and personalized risk weight can be tuned by the designer of the driver support system. For each driver type, a different parametrization is needed. Here, it is important to cover a wide range of driving types with the parametrization. In this paper, we set the scaling risk factor for the defensive driver to 
$\alpha_{\text{def}}\hspace{-0.03cm}=\hspace{-0.03cm}1$, for the normal driver to $\alpha_{\text{norm}}\hspace{-0.03cm}=\hspace{-0.03cm}0.5$ and for the confident driver to $\alpha_{\text{conf}}\hspace{-0.03cm}=\hspace{-0.03cm}0.04$. The risk weights for the different driver types were set to $w_{\text{def}} = 10$, $w_{\text{norm}} = 1$ and $w_{\text{conf}} = 0.1$ in this paper. 

\begin{table}[t!]
\fontsize{7.1}{8.1}\selectfont 
\vspace*{0.28cm}

\centering
\resizebox{0.91\columnwidth}{!}{
\begin{tabular}{c c c} 
\toprule
Parameter & Description & Value \\
\midrule
$v_{\text{off}}$ & velocity offset (forecast error) $[\unit[]{m/s}]$ & $-4$ \\
$a_{\text{intent}}$ & acceleration value (inference error) $[\unit[]{m/s^2}]$  & $-1.5$ \\
$t_{\text{intent}}$ & accceleration duration (inference error) $[\unit[]{s}]$ & $3$ \\
$\alpha_{\text{def}}$ & risk factor (defensive driver) & $1$ \\ 
$\alpha_{\text{norm}}$ & risk factor (normal driver) & $0.5$ \\
$\alpha_{\text{conf}}$ & risk factor (confident driver) & $0.04$ \\
$w_{\text{def}}$ & risk weight (defensive driver) & $10$ \\
$w_{\text{norm}}$ & risk weight (normal driver) & $1$ \\
$w_{\text{conf}}$ & risk weight (confident driver) & $0.1$ \\
$s_{\text{max}}$ & prediction horizon $[\unit[]{s}]$ & $8$ \\
\bottomrule
\end{tabular}
}
\vspace{0.05cm}
\caption[]{Parameter descriptions and values for the proposed human-based risk model of this paper. The parameter values were chosen based on empirical studies.}
\vspace{-0.1cm}
\label{tab:parameters}
\end{table}

In comparison, the baseline risk model \cite{puphal2019} does not use any human models and therefore does not consider the same set of parameters. The risk factor and risk weight for this risk model equals here the parametrization of the normal driver from the human-based risk model (i.e., $\alpha_{\text{norm}} = 0.5$ and $w_{\text{norm}} = 1$). Additionaly, we also set the same prediction horizon of $s_{\max}=\unit[8]{s}$ for the baseline risk model and the proposed human-based risk model of this paper.

%% file: images/5_human_based_risk_model.pdf_tex
\begingroup%
  \makeatletter%
  \providecommand\color[2][]{%
    \errmessage{(Inkscape) Color is used for the text in Inkscape, but the package 'color.sty' is not loaded}%
    \renewcommand\color[2][]{}%
  }%
  \providecommand\transparent[1]{%
    \errmessage{(Inkscape) Transparency is used (non-zero) for the text in Inkscape, but the package 'transparent.sty' is not loaded}%
    \renewcommand\transparent[1]{}%
  }%
  \providecommand\rotatebox[2]{#2}%
  \newcommand*\fsize{\dimexpr\f@size pt\relax}%
  \newcommand*\lineheight[1]{\fontsize{\fsize}{#1\fsize}\selectfont}%
  \ifx\svgwidth\undefined%
    \setlength{\unitlength}{310.1709137bp}%
    \ifx\svgscale\undefined%
      \relax%
    \else%
      \setlength{\unitlength}{\unitlength * \real{\svgscale}}%
    \fi%
  \else%
    \setlength{\unitlength}{\svgwidth}%
  \fi%
  \global\let\svgwidth\undefined%
  \global\let\svgscale\undefined%
  \makeatother%
  \begin{picture}(1,0.46047644)%
    \lineheight{1}%
    \setlength\tabcolsep{0pt}%
    \put(0,0){\includegraphics[width=\unitlength,page=1]{5_human_based_risk_model.pdf}}%
    \put(-0.00132475,0.29752922){\color[rgb]{0,0,0}\makebox(0,0)[lt]{\lineheight{1.25}\smash{\begin{tabular}[t]{l}driver errors\end{tabular}}}}%
    \put(-0.00504485,0.42453457){\color[rgb]{0,0,0}\makebox(0,0)[lt]{\lineheight{1.25}\smash{\begin{tabular}[t]{l}vehicle states\end{tabular}}}}%
    \put(0.43837063,0.45300964){\color[rgb]{0,0,0}\makebox(0,0)[lt]{\lineheight{1.25}\smash{\begin{tabular}[t]{l}driver\\behavior \end{tabular}}}}%
    \put(0.82819127,0.40729776){\color[rgb]{0,0,0}\makebox(0,0)[lt]{\lineheight{1.25}\smash{\begin{tabular}[t]{l}warning signal\end{tabular}}}}%
    \put(0.52976858,0.17564835){\color[rgb]{0,0,0}\makebox(0,0)[lt]{\lineheight{1.25}\smash{\begin{tabular}[t]{l}driver \\type\end{tabular}}}}%
    \put(0.16961642,0.17762071){\color[rgb]{0,0,0}\makebox(0,0)[lt]{\lineheight{1.25}\smash{\begin{tabular}[t]{l}-considers\\driver perception \end{tabular}}}}%
    \put(0,0){\includegraphics[width=\unitlength,page=2]{5_human_based_risk_model.pdf}}%
    \put(0.65415112,0.04371495){\color[rgb]{0,0,0}\makebox(0,0)[lt]{\lineheight{1.25}\smash{\begin{tabular}[t]{l}-considers\\driver personalization\end{tabular}}}}%
    \put(0.22324398,0.38434629){\color[rgb]{1,1,1}\makebox(0,0)[lt]{\lineheight{1.25}\smash{\begin{tabular}[t]{l}\textbf{Perceived}\end{tabular}}}}%
    \put(0,0){\includegraphics[width=\unitlength,page=3]{5_human_based_risk_model.pdf}}%
    \put(0.42470137,0.05245906){\color[rgb]{1,1,1}\makebox(0,0)[lt]{\lineheight{1.25}\smash{\begin{tabular}[t]{l}\textbf{Risk factor }\end{tabular}}}}%
    \put(0.22603129,0.34449415){\color[rgb]{1,1,1}\makebox(0,0)[lt]{\lineheight{1.25}\smash{\begin{tabular}[t]{l}\textbf{Risk Maps}\end{tabular}}}}%
    \put(0.61186279,0.35730258){\color[rgb]{1,1,1}\makebox(0,0)[lt]{\lineheight{1.25}\smash{\begin{tabular}[t]{l}\textbf{Risk model}\end{tabular}}}}%
  \end{picture}%
\endgroup%

%% file: chapters/experiments.tex
\section{Experiments}
\label{sec:experiments}

The following section describes the experiments that show driver support using the presented human-based risk model of this paper. In the experiments, simulations were done with a custom simulator written in python. The simulator can simulate vehicles as point masses without detailed vehicle dynamics, allowing for high-level testing of driver support systems. 
The human driver information of the given driver error and driver type can be set in the simulator and was assumed fix over one simulation. 

\begin{figure}[t!]
  \centering
  \vspace*{-0.03cm}
  \resizebox{0.94\linewidth}{!}{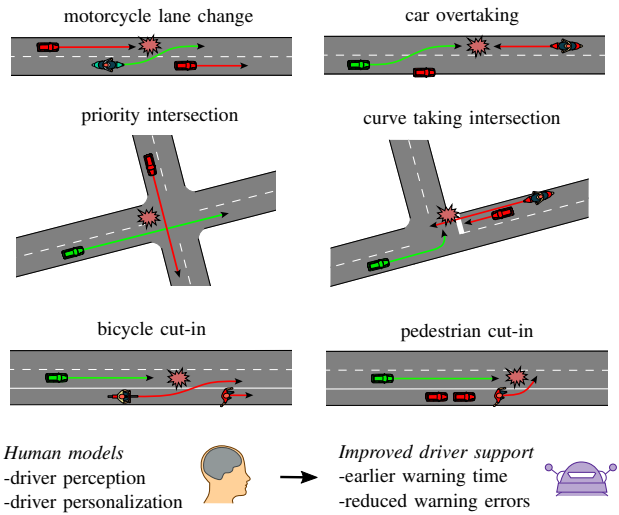}
  \vspace*{-0.01cm}
  \caption[]{Interactive driving scenarios which are analyzed in the experiments of this paper. The driver of the vehicle that is supported in the scenarios is colored green. We show that the proposed human-based risk model allows to improve driver support.}
  \label{fig:experiments}
  \vspace{-0.01cm}
\end{figure}

Overall, we simulated variations of six interactive driving scenarios. Besides the ``motorcycle lane change'' scenario, which was explained in the introduction, we analyze the scenarios of ``car overtaking'', ``priority intersection'', ``curve taking intersection'', ``bicycle cut-in'' and ``pedestrian cut-in''. The scenarios are depicted in Fig. \ref{fig:experiments}.
For each scenario, we varied the system inputs. For the experiments, we then compared our novel approach with the baseline risk model~\cite{puphal2019} that uses no human driver information.

In the next subsection, we will describe the scenario variation in detail. The simulation results are given afterwards and finally, we will conclude with a discussion of the limitations of the experiments and the proposed human-based risk model.

\subsection{Scenario Variation}
To simulate a wide range of driving conditions for the driving scenarios, we simulated each scenario with different driver errors and driver types for the ego driver that is supported by the system.
The variations for the driver errors were set either to no driver error for any other vehicle, or to a notice error with $\text{NE}(t) = 1$, forecast error with $\text{FE}(t) = 1$ or inference error with $\text{IE}(t) = 1$ for the most important other vehicle in the scenario. The variations for the driver type included a defensive driver, normal driver or confident driver. 

In addition, each driving scenario had three variations in which we varied the urgency of the driving scenario. We varied starting distance of the vehicles interacting, velocity of the interaction vehicles or timing of a behavior change in the driving scenario (i.e., timing of the lane change, overtaking or sudden cut-in). 

Regarding parametrization in the human-based risk model, the following guidelines were kept for the driver personalization. First, the defensive risk factor and risk weight were set so that no warning is issued in the results for no driver error, and an earlier warning is achieved when the driver has a driver error compared to the baseline risk model. Second, the confident risk factor and risk weight were set so that fewer false positives were achieved for no driver error, and there is on average no later warning when the driver has a driver error compared to the baseline. These guidelines need to be kept to fulfill a fair parametrization. This led to the parameter values listed in Tab. \ref{tab:parameters}.

\subsection{Simulation Results}
The results of the experiments for the human-based risk model show that an improvement of the driver support can be achieved compared to the baseline risk model.

\begin{figure}[t!]
  \centering
  \vspace{-0.06cm}
  \resizebox{0.96\linewidth}{!}{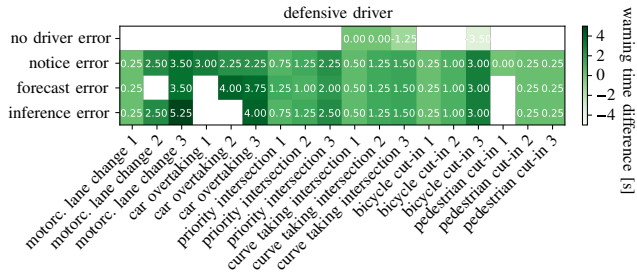}
  \vspace{0.1cm}
  \caption[]{The image shows the warning time improvement of the human-based risk model for a defensive driver for different variations of the analyzed interactive driving scenarios. The baseline risk model \cite{puphal2019} does not consider human driver information. The highest warning time improvement is achieved for the shown defensive driver.}
  \vspace{-0.15cm}
  \label{fig:time_defensive}
\end{figure}

\begin{figure}[t!]
  \centering
  \vspace*{-0.22cm}
  \resizebox{0.96\linewidth}{!}{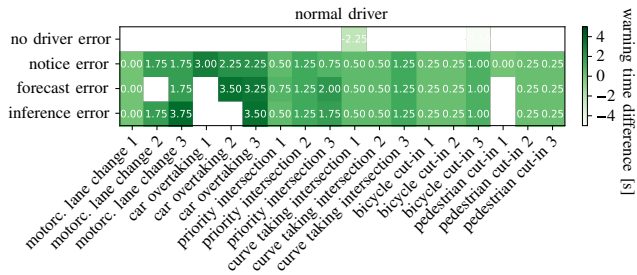}
  \vspace{0.05cm}
  \caption[]{Warning time improvement for normal driver for the same driving scenario variations.}
  \vspace{-0.1cm}
  \label{fig:time_normal}
\end{figure}

\begin{figure}[t!]
  \centering
  \vspace*{-0.3cm}
  \resizebox{0.96\linewidth}{!}{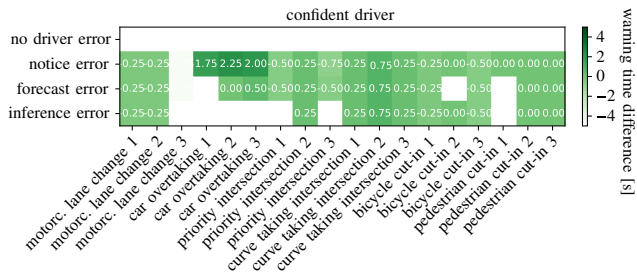}
  \vspace{0.05cm}
  \caption[]{Warning time improvement for confident driver for the same driving scenario variations.}
  \vspace{-0.1cm}
  \label{fig:time_confident}
\end{figure}

The warning time improvement for a defensive driver is shown in Fig. \ref{fig:time_defensive}, whereby the improvement for a normal driver and the improvement for a confident driver is depicted in Fig. \ref{fig:time_normal} and Fig. \ref{fig:time_confident}, respectively. The figures show heatmaps that denote the warning time difference between the proposed human-based risk model and the baseline risk model for the simulation of each scenario variation. The abscissa in the heatmaps shows the scenario variation name and the ordinate shows the driver error which was simulated. The warning time improvements reach up to $5.25$ seconds when the driver has a driver error.

As can be seen, the scenarios of the motorcycle lane change (i.e., ``motorc. lane change 1 - 3'') and car overtaking (i.e., ``car overtaking 1 - 3'') are the easiest to handle for the human-based risk model. Larger warning time improvements are achieved in all figures because the risk increases slowly over the time in these scenarios and there is more potential to warn earlier. In contrast, the pedestrian cut-in scenario (i.e., ``pedestrian cut-in 1 - 3'') is challenging for the proposed risk model. Smaller warning time improvements are achieved in the figures because the risk increases very sudden and there is not much time to warn in advance. 

For the warning time improvement, Fig. \ref{fig:time_defensive} shows that the highest improvement can be achieved for the defensive driver. The driver can be warned on average $1.55$ seconds earlier when the driver has a driver error. 

\begin{figure}[t!]
  \centering
  \vspace{-0.07cm}
  \resizebox{0.96\linewidth}{!}{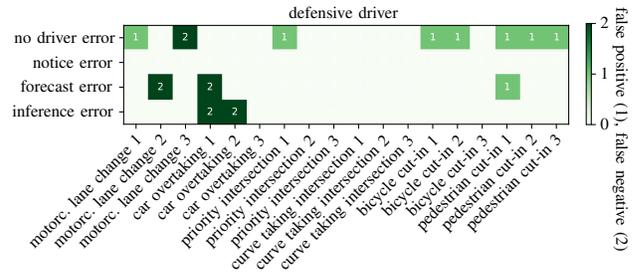}
  \vspace{0.1cm}
  \caption[]{The image shows the warning error reduction of the human-based risk model for a defensive driver in comparison to the baseline risk model. The baseline risk model does not use human driver information.} 
  \label{fig:error_defensive}
\end{figure}

\begin{figure}[t!]
  \centering
  \vspace*{-0.33cm}
  \resizebox{0.96\linewidth}{!}{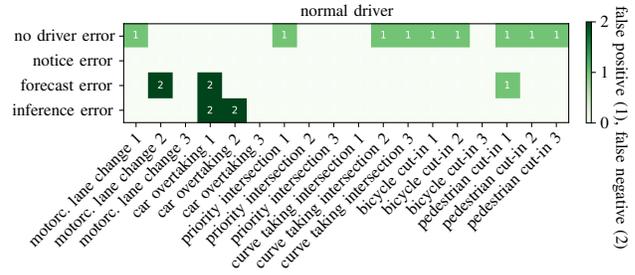}
  \vspace{0.05cm}
  \caption[]{Warning error reduction for normal driver for the same variations.}
  \label{fig:error_normal}
\end{figure}

\begin{figure}[t!]
  \centering
  \vspace*{-0.3cm}
  \resizebox{0.96\linewidth}{!}{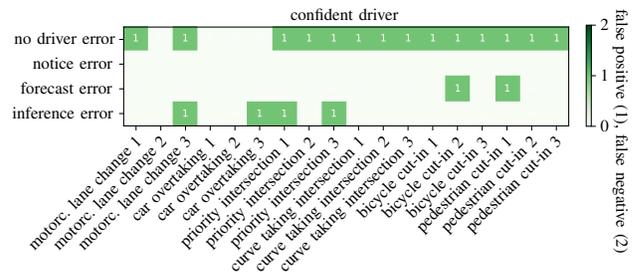}
  \vspace{0.05cm}
  \caption[]{Warning error reduction for confident driver for the same driving scenario variations. In the confident driver case, false positives warning errors are reduced the most.}
  \vspace{-0.05cm}
  \label{fig:error_confident}
\end{figure}

On the same line, the warning error reduction  improvements for a defensive driver, a normal driver and a confident driver are shown in Fig. \ref{fig:error_defensive}, Fig. \ref{fig:error_normal} and Fig. \ref{fig:error_confident}. The same heatmaps are given in the figures, this time showing false positive and false negative warning error reductions for each driving scenario variation. While no warning error reductions are denoted with the value $0$, false positive error and false negative error reductions are denoted with a value of $1$ and $2$, respectively. 

As can be seen in the figures, human driver information helps in our simulation results more in reducing false positive warning errors than in reducing false negative errors. However, false negative warning errors can also be reduced with the proposed human-based risk model. The highest improvement is achieved for a confident driver. Here, false positive errors can be reduced for 28 $\%$ of the simulated driving scenario variations (compare the many green entries within the heatmap in Fig. \ref{fig:error_confident}).

Tab. \ref{tab:Statistics} finally shows the statistical details of the results. The table depicts absolute value, variance, minimum and maximum of the warning time improvement and shows false positive (FP) and false negative error reductions (FN) in percent for the defensive, normal and confident driver cases. The table highlights the highest warning time improvement for the defensive driver with $1.55$ seconds and the highest warning error reduction improvement for the confident driver with 28$~\%$ in bold. The normal driver has moreover the least variance in the warning time improvement with a value of $1.02$ seconds for the human-based risk model.

\begin{table}[t!]
\fontsize{9.15}{12}\selectfont 
\vspace{0.15cm}
\centering
\resizebox{0.91\columnwidth}{!}{
\begin{tabular}{|c|p{0.89cm}|p{0.89cm}|p{0.89cm}|p{0.93cm}|p{0.89cm}|p{0.93cm}|p{0.74cm}|} 
\cline{2-7}
\multicolumn{1}{c|}{} & \multicolumn{4}{c|}{Warning time improvement} & \multicolumn{2}{c|}{Error reduction} \\
\cline{2-7}
\multicolumn{1}{c|}{} & abs $[\unit[]{s}]$ & var $[\unit[]{s}]$ & min $[\unit[]{s}]$ & max $[\unit[]{s}]$ & FP $[\unit[]{\%}]$ & FN $[\unit[]{\%}]$ \\
\hline
defensive & \textbf{1.55} & 1.67 & 0.0 & 5.25 & 11 & 7 \\
\hline
normal & 1.07 & \textbf{1.02} & 0.0 & 3.75 & 14 & 6 \\
\hline
confident & -0.24 & 3.07 & -7.75 & 2.25 & \textbf{28} & 0 \\
\hline
\end{tabular}
}
\vspace{-0.03cm}
\caption[]{Statistics of warning time improvement and error reduction from the human-based risk model compared to the baseline risk model not using human driver information. Notable results are shown in bold. In total, we ran over 200 different simulations. The warning time statistics is only shown for the cases when the driver has a driver error.}
\label{tab:Statistics}
\end{table}

\subsection{Discussion}
\makebox[0.96\linewidth][s]{The experiments show that the proposed human-based} 

\noindent risk model of this paper is able to improve driver support and generalizes to the different driving scenarios used in this paper. In this last section, we will however discuss the limitations of the experiments and explain possible improvements for future work.

In the experiments, the driver state estimation module was assumed to be given. Hence, there were no errors and noise in the human driver information regarding driver error and risk factor. 

As an example, the personalization may differ between the assumed risk factor and the actual driver preference. 
The error in personalization should not exceed $\Delta \alpha=\alpha_{\text{pers}}-\alpha_{\text{normal}}$, which results in $\Delta \alpha \hspace{-0.01cm} \approx \hspace{-0.01cm} 0.5$ for the given parametrization. Otherwise, the improvement in the human-based risk model is not observable. A user study should be done to investigate this simulation-to-reality gap. 

Furthermore, the variation of the driving scenarios was manually chosen by the authors. We varied the scenarios by always sampling one hard, one medium difficult and one easy driving variation for the human-based risk model according to the urgency of the potential collision in the scenario. However, a fair balance of driving conditions cannot be ensured accurately. For example, an oversampling of sudden cut-in driving situations would lead to less average warning time improvements. With these limiations in mind, the experiments show a promising trend that human factors allows for effective and improved driver support.

%% file: images/1_driving_scenarios7.pdf_tex
\begingroup%
  \makeatletter%
  \providecommand\color[2][]{%
    \errmessage{(Inkscape) Color is used for the text in Inkscape, but the package 'color.sty' is not loaded}%
    \renewcommand\color[2][]{}%
  }%
  \providecommand\transparent[1]{%
    \errmessage{(Inkscape) Transparency is used (non-zero) for the text in Inkscape, but the package 'transparent.sty' is not loaded}%
    \renewcommand\transparent[1]{}%
  }%
  \providecommand\rotatebox[2]{#2}%
  \newcommand*\fsize{\dimexpr\f@size pt\relax}%
  \newcommand*\lineheight[1]{\fontsize{\fsize}{#1\fsize}\selectfont}%
  \ifx\svgwidth\undefined%
    \setlength{\unitlength}{310.84406335bp}%
    \ifx\svgscale\undefined%
      \relax%
    \else%
      \setlength{\unitlength}{\unitlength * \real{\svgscale}}%
    \fi%
  \else%
    \setlength{\unitlength}{\svgwidth}%
  \fi%
  \global\let\svgwidth\undefined%
  \global\let\svgscale\undefined%
  \makeatother%
  \begin{picture}(1,0.86402043)%
    \lineheight{1}%
    \setlength\tabcolsep{0pt}%
    \put(0,0){\includegraphics[width=\unitlength,page=1]{1_driving_scenarios7.pdf}}%
    \put(0.58901164,-1.10053929){\color[rgb]{0,0,0}\makebox(0,0)[lt]{\begin{minipage}{0.23366356\unitlength}\end{minipage}}}%
    \put(0,0){\includegraphics[width=\unitlength,page=2]{1_driving_scenarios7.pdf}}%
    \put(0.58901164,-1.10053929){\color[rgb]{0,0,0}\makebox(0,0)[lt]{\begin{minipage}{0.23366356\unitlength}\end{minipage}}}%
    \put(0.09327789,0.80896892){\color[rgb]{0,0,0}\makebox(0,0)[lt]{\lineheight{1.25}\smash{\begin{tabular}[t]{l}motorcycle lane change\end{tabular}}}}%
    \put(0.65268007,0.8110812){\color[rgb]{0,0,0}\makebox(0,0)[lt]{\lineheight{1.25}\smash{\begin{tabular}[t]{l}car overtaking\end{tabular}}}}%
    \put(0.12229935,0.64690551){\color[rgb]{0,0,0}\makebox(0,0)[lt]{\lineheight{1.25}\smash{\begin{tabular}[t]{l}priority intersection\end{tabular}}}}%
    \put(0.58392765,0.64139328){\color[rgb]{0,0,0}\makebox(0,0)[lt]{\lineheight{1.25}\smash{\begin{tabular}[t]{l}curve taking intersection\end{tabular}}}}%
    \put(0.14819958,0.29683435){\color[rgb]{0,0,0}\makebox(0,0)[lt]{\lineheight{1.25}\smash{\begin{tabular}[t]{l}bicycle cut-in\end{tabular}}}}%
    \put(0.64484824,0.29362873){\color[rgb]{0,0,0}\makebox(0,0)[lt]{\lineheight{1.25}\smash{\begin{tabular}[t]{l}pedestrian cut-in\end{tabular}}}}%
    \put(0,0){\includegraphics[width=\unitlength,page=3]{1_driving_scenarios7.pdf}}%
    \put(-0.00487948,0.0926762){\color[rgb]{0,0,0}\makebox(0,0)[lt]{\lineheight{1.25}\smash{\begin{tabular}[t]{l}\textit{Human models}\\-driver perception\\-driver personalization\end{tabular}}}}%
    \put(0.54343228,0.09501495){\color[rgb]{0,0,0}\makebox(0,0)[lt]{\lineheight{1.25}\smash{\begin{tabular}[t]{l}\textit{Improved driver support}\\-earlier warning time\\-reduced warning errors\end{tabular}}}}%
  \end{picture}%
\endgroup%

%% file: images/experiments/time_improvement_defensive9.pdf_tex
\begingroup%
  \makeatletter%
  \providecommand\color[2][]{%
    \errmessage{(Inkscape) Color is used for the text in Inkscape, but the package 'color.sty' is not loaded}%
    \renewcommand\color[2][]{}%
  }%
  \providecommand\transparent[1]{%
    \errmessage{(Inkscape) Transparency is used (non-zero) for the text in Inkscape, but the package 'transparent.sty' is not loaded}%
    \renewcommand\transparent[1]{}%
  }%
  \providecommand\rotatebox[2]{#2}%
  \newcommand*\fsize{\dimexpr\f@size pt\relax}%
  \newcommand*\lineheight[1]{\fontsize{\fsize}{#1\fsize}\selectfont}%
  \ifx\svgwidth\undefined%
    \setlength{\unitlength}{370.10212191bp}%
    \ifx\svgscale\undefined%
      \relax%
    \else%
      \setlength{\unitlength}{\unitlength * \real{\svgscale}}%
    \fi%
  \else%
    \setlength{\unitlength}{\svgwidth}%
  \fi%
  \global\let\svgwidth\undefined%
  \global\let\svgscale\undefined%
  \makeatother%
  \begin{picture}(1,0.44235964)%
    \lineheight{1}%
    \setlength\tabcolsep{0pt}%
    \put(0,0){\includegraphics[width=\unitlength,page=1]{time_improvement_defensive9.pdf}}%
    \put(0.03367651,0.02289768){\color[rgb]{0,0,0}\rotatebox{45.00000162}{\makebox(0,0)[lt]{\lineheight{1.25}\smash{\begin{tabular}[t]{l}motorc. lane change 1\end{tabular}}}}}%
    \put(0,0){\includegraphics[width=\unitlength,page=2]{time_improvement_defensive9.pdf}}%
    \put(0.07332085,0.02289768){\color[rgb]{0,0,0}\rotatebox{45.00000162}{\makebox(0,0)[lt]{\lineheight{1.25}\smash{\begin{tabular}[t]{l}motorc. lane change 2\end{tabular}}}}}%
    \put(0,0){\includegraphics[width=\unitlength,page=3]{time_improvement_defensive9.pdf}}%
    \put(0.1129652,0.02289769){\color[rgb]{0,0,0}\rotatebox{45.00000162}{\makebox(0,0)[lt]{\lineheight{1.25}\smash{\begin{tabular}[t]{l}motorc. lane change 3\end{tabular}}}}}%
    \put(0,0){\includegraphics[width=\unitlength,page=4]{time_improvement_defensive9.pdf}}%
    \put(0.19877112,0.06840386){\color[rgb]{0,0,0}\rotatebox{45.00000162}{\makebox(0,0)[lt]{\lineheight{1.25}\smash{\begin{tabular}[t]{l}car overtaking 1\end{tabular}}}}}%
    \put(0,0){\includegraphics[width=\unitlength,page=5]{time_improvement_defensive9.pdf}}%
    \put(0.23841546,0.06840386){\color[rgb]{0,0,0}\rotatebox{45.00000162}{\makebox(0,0)[lt]{\lineheight{1.25}\smash{\begin{tabular}[t]{l}car overtaking 2\end{tabular}}}}}%
    \put(0,0){\includegraphics[width=\unitlength,page=6]{time_improvement_defensive9.pdf}}%
    \put(0.2780598,0.06840386){\color[rgb]{0,0,0}\rotatebox{45.00000162}{\makebox(0,0)[lt]{\lineheight{1.25}\smash{\begin{tabular}[t]{l}car overtaking 3\end{tabular}}}}}%
    \put(0,0){\includegraphics[width=\unitlength,page=7]{time_improvement_defensive9.pdf}}%
    \put(0.27673091,0.02677524){\color[rgb]{0,0,0}\rotatebox{45.00000162}{\makebox(0,0)[lt]{\lineheight{1.25}\smash{\begin{tabular}[t]{l}priority intersection 1\end{tabular}}}}}%
    \put(0,0){\includegraphics[width=\unitlength,page=8]{time_improvement_defensive9.pdf}}%
    \put(0.31637525,0.02677523){\color[rgb]{0,0,0}\rotatebox{45.00000162}{\makebox(0,0)[lt]{\lineheight{1.25}\smash{\begin{tabular}[t]{l}priority intersection 2\end{tabular}}}}}%
    \put(0,0){\includegraphics[width=\unitlength,page=9]{time_improvement_defensive9.pdf}}%
    \put(0.35601959,0.02677523){\color[rgb]{0,0,0}\rotatebox{45.00000162}{\makebox(0,0)[lt]{\lineheight{1.25}\smash{\begin{tabular}[t]{l}priority intersection 3\end{tabular}}}}}%
    \put(0,0){\includegraphics[width=\unitlength,page=10]{time_improvement_defensive9.pdf}}%
    \put(0.35680702,-0.01208167){\color[rgb]{0,0,0}\rotatebox{45.00000162}{\makebox(0,0)[lt]{\lineheight{1.25}\smash{\begin{tabular}[t]{l}curve taking intersection 1\end{tabular}}}}}%
    \put(0,0){\includegraphics[width=\unitlength,page=11]{time_improvement_defensive9.pdf}}%
    \put(0.39645137,-0.01208166){\color[rgb]{0,0,0}\rotatebox{45.00000162}{\makebox(0,0)[lt]{\lineheight{1.25}\smash{\begin{tabular}[t]{l}curve taking intersection 2\end{tabular}}}}}%
    \put(0,0){\includegraphics[width=\unitlength,page=12]{time_improvement_defensive9.pdf}}%
    \put(0.43609571,-0.01208167){\color[rgb]{0,0,0}\rotatebox{45.00000162}{\makebox(0,0)[lt]{\lineheight{1.25}\smash{\begin{tabular}[t]{l}curve taking intersection 3\end{tabular}}}}}%
    \put(0,0){\includegraphics[width=\unitlength,page=13]{time_improvement_defensive9.pdf}}%
    \put(0.56158347,0.07376176){\color[rgb]{0,0,0}\rotatebox{45.00000162}{\makebox(0,0)[lt]{\lineheight{1.25}\smash{\begin{tabular}[t]{l}bicycle cut-in 1\end{tabular}}}}}%
    \put(0,0){\includegraphics[width=\unitlength,page=14]{time_improvement_defensive9.pdf}}%
    \put(0.60122781,0.07376176){\color[rgb]{0,0,0}\rotatebox{45.00000162}{\makebox(0,0)[lt]{\lineheight{1.25}\smash{\begin{tabular}[t]{l}bicycle cut-in 2\end{tabular}}}}}%
    \put(0,0){\includegraphics[width=\unitlength,page=15]{time_improvement_defensive9.pdf}}%
    \put(0.64087215,0.07376176){\color[rgb]{0,0,0}\rotatebox{45.00000162}{\makebox(0,0)[lt]{\lineheight{1.25}\smash{\begin{tabular}[t]{l}bicycle cut-in 3\end{tabular}}}}}%
    \put(0,0){\includegraphics[width=\unitlength,page=16]{time_improvement_defensive9.pdf}}%
    \put(0.65592446,0.05179129){\color[rgb]{0,0,0}\rotatebox{45.00000162}{\makebox(0,0)[lt]{\lineheight{1.25}\smash{\begin{tabular}[t]{l}pedestrian cut-in 1\end{tabular}}}}}%
    \put(0,0){\includegraphics[width=\unitlength,page=17]{time_improvement_defensive9.pdf}}%
    \put(0.6955688,0.05179129){\color[rgb]{0,0,0}\rotatebox{45.00000162}{\makebox(0,0)[lt]{\lineheight{1.25}\smash{\begin{tabular}[t]{l}pedestrian cut-in 2\end{tabular}}}}}%
    \put(0,0){\includegraphics[width=\unitlength,page=18]{time_improvement_defensive9.pdf}}%
    \put(0.73521315,0.0517913){\color[rgb]{0,0,0}\rotatebox{45.00000162}{\makebox(0,0)[lt]{\lineheight{1.25}\smash{\begin{tabular}[t]{l}pedestrian cut-in 3\end{tabular}}}}}%
    \put(0,0){\includegraphics[width=\unitlength,page=19]{time_improvement_defensive9.pdf}}%
    \put(0.0016462,0.35383829){\color[rgb]{0,0,0}\makebox(0,0)[lt]{\lineheight{1.25}\smash{\begin{tabular}[t]{l}no driver error\end{tabular}}}}%
    \put(0,0){\includegraphics[width=\unitlength,page=20]{time_improvement_defensive9.pdf}}%
    \put(0.03126569,0.31419397){\color[rgb]{0,0,0}\makebox(0,0)[lt]{\lineheight{1.25}\smash{\begin{tabular}[t]{l}notice error\end{tabular}}}}%
    \put(0,0){\includegraphics[width=\unitlength,page=21]{time_improvement_defensive9.pdf}}%
    \put(0.0113245,0.27454962){\color[rgb]{0,0,0}\makebox(0,0)[lt]{\lineheight{1.25}\smash{\begin{tabular}[t]{l}forecast error\end{tabular}}}}%
    \put(0,0){\includegraphics[width=\unitlength,page=22]{time_improvement_defensive9.pdf}}%
    \put(-0.0019588,0.23490527){\color[rgb]{0,0,0}\makebox(0,0)[lt]{\lineheight{1.25}\smash{\begin{tabular}[t]{l}inference error\end{tabular}}}}%
    \put(0,0){\includegraphics[width=\unitlength,page=23]{time_improvement_defensive9.pdf}}%
    \put(0.4394182,0.39402388){\color[rgb]{0,0,0}\makebox(0,0)[lt]{\lineheight{1.25}\smash{\begin{tabular}[t]{l}defensive driver\end{tabular}}}}%
    \put(0,0){\includegraphics[width=\unitlength,page=24]{time_improvement_defensive9.pdf}}%
    \put(0.95836384,0.23091084){\color[rgb]{0,0,0}\makebox(0,0)[lt]{\lineheight{1.25}\smash{\begin{tabular}[t]{l}4\end{tabular}}}}%
    \put(0,0){\includegraphics[width=\unitlength,page=25]{time_improvement_defensive9.pdf}}%
    \put(0.95836384,0.26264131){\color[rgb]{0,0,0}\makebox(0,0)[lt]{\lineheight{1.25}\smash{\begin{tabular}[t]{l}2\end{tabular}}}}%
    \put(0,0){\includegraphics[width=\unitlength,page=26]{time_improvement_defensive9.pdf}}%
    \put(0.94094777,0.29437179){\color[rgb]{0,0,0}\makebox(0,0)[lt]{\lineheight{1.25}\smash{\begin{tabular}[t]{l}0\end{tabular}}}}%
    \put(0,0){\includegraphics[width=\unitlength,page=27]{time_improvement_defensive9.pdf}}%
    \put(0.94094777,0.32610227){\color[rgb]{0,0,0}\makebox(0,0)[lt]{\lineheight{1.25}\smash{\begin{tabular}[t]{l}2\end{tabular}}}}%
    \put(0,0){\includegraphics[width=\unitlength,page=28]{time_improvement_defensive9.pdf}}%
    \put(0.94094777,0.35783275){\color[rgb]{0,0,0}\makebox(0,0)[lt]{\lineheight{1.25}\smash{\begin{tabular}[t]{l}4\end{tabular}}}}%
    \put(0.98420782,0.41205406){\color[rgb]{0,0,0}\rotatebox{-90}{\makebox(0,0)[lt]{\lineheight{1.25}\smash{\begin{tabular}[t]{l}warning time difference [s]\end{tabular}}}}}%
    \put(0,0){\includegraphics[width=\unitlength,page=29]{time_improvement_defensive9.pdf}}%
  \end{picture}%
\endgroup%

%% file: images/experiments/time_improvement_normal1.pdf_tex
\begingroup%
  \makeatletter%
  \providecommand\color[2][]{%
    \errmessage{(Inkscape) Color is used for the text in Inkscape, but the package 'color.sty' is not loaded}%
    \renewcommand\color[2][]{}%
  }%
  \providecommand\transparent[1]{%
    \errmessage{(Inkscape) Transparency is used (non-zero) for the text in Inkscape, but the package 'transparent.sty' is not loaded}%
    \renewcommand\transparent[1]{}%
  }%
  \providecommand\rotatebox[2]{#2}%
  \newcommand*\fsize{\dimexpr\f@size pt\relax}%
  \newcommand*\lineheight[1]{\fontsize{\fsize}{#1\fsize}\selectfont}%
  \ifx\svgwidth\undefined%
    \setlength{\unitlength}{370.10212191bp}%
    \ifx\svgscale\undefined%
      \relax%
    \else%
      \setlength{\unitlength}{\unitlength * \real{\svgscale}}%
    \fi%
  \else%
    \setlength{\unitlength}{\svgwidth}%
  \fi%
  \global\let\svgwidth\undefined%
  \global\let\svgscale\undefined%
  \makeatother%
  \begin{picture}(1,0.44235964)%
    \lineheight{1}%
    \setlength\tabcolsep{0pt}%
    \put(0,0){\includegraphics[width=\unitlength,page=1]{time_improvement_normal1.pdf}}%
    \put(0.03367651,0.02289768){\color[rgb]{0,0,0}\rotatebox{45.00000162}{\makebox(0,0)[lt]{\lineheight{1.25}\smash{\begin{tabular}[t]{l}motorc. lane change 1\end{tabular}}}}}%
    \put(0,0){\includegraphics[width=\unitlength,page=2]{time_improvement_normal1.pdf}}%
    \put(0.07332085,0.02289768){\color[rgb]{0,0,0}\rotatebox{45.00000162}{\makebox(0,0)[lt]{\lineheight{1.25}\smash{\begin{tabular}[t]{l}motorc. lane change 2\end{tabular}}}}}%
    \put(0,0){\includegraphics[width=\unitlength,page=3]{time_improvement_normal1.pdf}}%
    \put(0.1129652,0.02289769){\color[rgb]{0,0,0}\rotatebox{45.00000162}{\makebox(0,0)[lt]{\lineheight{1.25}\smash{\begin{tabular}[t]{l}motorc. lane change 3\end{tabular}}}}}%
    \put(0,0){\includegraphics[width=\unitlength,page=4]{time_improvement_normal1.pdf}}%
    \put(0.19877112,0.06840386){\color[rgb]{0,0,0}\rotatebox{45.00000162}{\makebox(0,0)[lt]{\lineheight{1.25}\smash{\begin{tabular}[t]{l}car overtaking 1\end{tabular}}}}}%
    \put(0,0){\includegraphics[width=\unitlength,page=5]{time_improvement_normal1.pdf}}%
    \put(0.23841546,0.06840386){\color[rgb]{0,0,0}\rotatebox{45.00000162}{\makebox(0,0)[lt]{\lineheight{1.25}\smash{\begin{tabular}[t]{l}car overtaking 2\end{tabular}}}}}%
    \put(0,0){\includegraphics[width=\unitlength,page=6]{time_improvement_normal1.pdf}}%
    \put(0.2780598,0.06840386){\color[rgb]{0,0,0}\rotatebox{45.00000162}{\makebox(0,0)[lt]{\lineheight{1.25}\smash{\begin{tabular}[t]{l}car overtaking 3\end{tabular}}}}}%
    \put(0,0){\includegraphics[width=\unitlength,page=7]{time_improvement_normal1.pdf}}%
    \put(0.27673091,0.02677524){\color[rgb]{0,0,0}\rotatebox{45.00000162}{\makebox(0,0)[lt]{\lineheight{1.25}\smash{\begin{tabular}[t]{l}priority intersection 1\end{tabular}}}}}%
    \put(0,0){\includegraphics[width=\unitlength,page=8]{time_improvement_normal1.pdf}}%
    \put(0.31637525,0.02677523){\color[rgb]{0,0,0}\rotatebox{45.00000162}{\makebox(0,0)[lt]{\lineheight{1.25}\smash{\begin{tabular}[t]{l}priority intersection 2\end{tabular}}}}}%
    \put(0,0){\includegraphics[width=\unitlength,page=9]{time_improvement_normal1.pdf}}%
    \put(0.35601959,0.02677523){\color[rgb]{0,0,0}\rotatebox{45.00000162}{\makebox(0,0)[lt]{\lineheight{1.25}\smash{\begin{tabular}[t]{l}priority intersection 3\end{tabular}}}}}%
    \put(0,0){\includegraphics[width=\unitlength,page=10]{time_improvement_normal1.pdf}}%
    \put(0.35680702,-0.01208167){\color[rgb]{0,0,0}\rotatebox{45.00000162}{\makebox(0,0)[lt]{\lineheight{1.25}\smash{\begin{tabular}[t]{l}curve taking intersection 1\end{tabular}}}}}%
    \put(0,0){\includegraphics[width=\unitlength,page=11]{time_improvement_normal1.pdf}}%
    \put(0.39645137,-0.01208166){\color[rgb]{0,0,0}\rotatebox{45.00000162}{\makebox(0,0)[lt]{\lineheight{1.25}\smash{\begin{tabular}[t]{l}curve taking intersection 2\end{tabular}}}}}%
    \put(0,0){\includegraphics[width=\unitlength,page=12]{time_improvement_normal1.pdf}}%
    \put(0.43609571,-0.01208167){\color[rgb]{0,0,0}\rotatebox{45.00000162}{\makebox(0,0)[lt]{\lineheight{1.25}\smash{\begin{tabular}[t]{l}curve taking intersection 3\end{tabular}}}}}%
    \put(0,0){\includegraphics[width=\unitlength,page=13]{time_improvement_normal1.pdf}}%
    \put(0.56158347,0.07376176){\color[rgb]{0,0,0}\rotatebox{45.00000162}{\makebox(0,0)[lt]{\lineheight{1.25}\smash{\begin{tabular}[t]{l}bicycle cut-in 1\end{tabular}}}}}%
    \put(0,0){\includegraphics[width=\unitlength,page=14]{time_improvement_normal1.pdf}}%
    \put(0.60122781,0.07376176){\color[rgb]{0,0,0}\rotatebox{45.00000162}{\makebox(0,0)[lt]{\lineheight{1.25}\smash{\begin{tabular}[t]{l}bicycle cut-in 2\end{tabular}}}}}%
    \put(0,0){\includegraphics[width=\unitlength,page=15]{time_improvement_normal1.pdf}}%
    \put(0.64087215,0.07376176){\color[rgb]{0,0,0}\rotatebox{45.00000162}{\makebox(0,0)[lt]{\lineheight{1.25}\smash{\begin{tabular}[t]{l}bicycle cut-in 3\end{tabular}}}}}%
    \put(0,0){\includegraphics[width=\unitlength,page=16]{time_improvement_normal1.pdf}}%
    \put(0.65592446,0.05179129){\color[rgb]{0,0,0}\rotatebox{45.00000162}{\makebox(0,0)[lt]{\lineheight{1.25}\smash{\begin{tabular}[t]{l}pedestrian cut-in 1\end{tabular}}}}}%
    \put(0,0){\includegraphics[width=\unitlength,page=17]{time_improvement_normal1.pdf}}%
    \put(0.6955688,0.05179129){\color[rgb]{0,0,0}\rotatebox{45.00000162}{\makebox(0,0)[lt]{\lineheight{1.25}\smash{\begin{tabular}[t]{l}pedestrian cut-in 2\end{tabular}}}}}%
    \put(0,0){\includegraphics[width=\unitlength,page=18]{time_improvement_normal1.pdf}}%
    \put(0.73521315,0.0517913){\color[rgb]{0,0,0}\rotatebox{45.00000162}{\makebox(0,0)[lt]{\lineheight{1.25}\smash{\begin{tabular}[t]{l}pedestrian cut-in 3\end{tabular}}}}}%
    \put(0,0){\includegraphics[width=\unitlength,page=19]{time_improvement_normal1.pdf}}%
    \put(0.0016462,0.35383829){\color[rgb]{0,0,0}\makebox(0,0)[lt]{\lineheight{1.25}\smash{\begin{tabular}[t]{l}no driver error\end{tabular}}}}%
    \put(0,0){\includegraphics[width=\unitlength,page=20]{time_improvement_normal1.pdf}}%
    \put(0.03126569,0.31419397){\color[rgb]{0,0,0}\makebox(0,0)[lt]{\lineheight{1.25}\smash{\begin{tabular}[t]{l}notice error\end{tabular}}}}%
    \put(0,0){\includegraphics[width=\unitlength,page=21]{time_improvement_normal1.pdf}}%
    \put(0.0113245,0.27454962){\color[rgb]{0,0,0}\makebox(0,0)[lt]{\lineheight{1.25}\smash{\begin{tabular}[t]{l}forecast error\end{tabular}}}}%
    \put(0,0){\includegraphics[width=\unitlength,page=22]{time_improvement_normal1.pdf}}%
    \put(-0.0019588,0.23490527){\color[rgb]{0,0,0}\makebox(0,0)[lt]{\lineheight{1.25}\smash{\begin{tabular}[t]{l}inference error\end{tabular}}}}%
    \put(0,0){\includegraphics[width=\unitlength,page=23]{time_improvement_normal1.pdf}}%
    \put(0.95836384,0.23091084){\color[rgb]{0,0,0}\makebox(0,0)[lt]{\lineheight{1.25}\smash{\begin{tabular}[t]{l}4\end{tabular}}}}%
    \put(0,0){\includegraphics[width=\unitlength,page=24]{time_improvement_normal1.pdf}}%
    \put(0.95836384,0.26264131){\color[rgb]{0,0,0}\makebox(0,0)[lt]{\lineheight{1.25}\smash{\begin{tabular}[t]{l}2\end{tabular}}}}%
    \put(0,0){\includegraphics[width=\unitlength,page=25]{time_improvement_normal1.pdf}}%
    \put(0.94094777,0.29437179){\color[rgb]{0,0,0}\makebox(0,0)[lt]{\lineheight{1.25}\smash{\begin{tabular}[t]{l}0\end{tabular}}}}%
    \put(0,0){\includegraphics[width=\unitlength,page=26]{time_improvement_normal1.pdf}}%
    \put(0.94094777,0.32610227){\color[rgb]{0,0,0}\makebox(0,0)[lt]{\lineheight{1.25}\smash{\begin{tabular}[t]{l}2\end{tabular}}}}%
    \put(0,0){\includegraphics[width=\unitlength,page=27]{time_improvement_normal1.pdf}}%
    \put(0.94094777,0.35783275){\color[rgb]{0,0,0}\makebox(0,0)[lt]{\lineheight{1.25}\smash{\begin{tabular}[t]{l}4\end{tabular}}}}%
    \put(0.98420782,0.41205406){\color[rgb]{0,0,0}\rotatebox{-90}{\makebox(0,0)[lt]{\lineheight{1.25}\smash{\begin{tabular}[t]{l}warning time difference [s]\end{tabular}}}}}%
    \put(0,0){\includegraphics[width=\unitlength,page=28]{time_improvement_normal1.pdf}}%
    \put(0.45766531,0.39424366){\color[rgb]{0,0,0}\makebox(0,0)[lt]{\lineheight{1.25}\smash{\begin{tabular}[t]{l}normal driver\end{tabular}}}}%
  \end{picture}%
\endgroup%

%% file: images/experiments/time_improvement_confident1.pdf_tex
\begingroup%
  \makeatletter%
  \providecommand\color[2][]{%
    \errmessage{(Inkscape) Color is used for the text in Inkscape, but the package 'color.sty' is not loaded}%
    \renewcommand\color[2][]{}%
  }%
  \providecommand\transparent[1]{%
    \errmessage{(Inkscape) Transparency is used (non-zero) for the text in Inkscape, but the package 'transparent.sty' is not loaded}%
    \renewcommand\transparent[1]{}%
  }%
  \providecommand\rotatebox[2]{#2}%
  \newcommand*\fsize{\dimexpr\f@size pt\relax}%
  \newcommand*\lineheight[1]{\fontsize{\fsize}{#1\fsize}\selectfont}%
  \ifx\svgwidth\undefined%
    \setlength{\unitlength}{370.10212191bp}%
    \ifx\svgscale\undefined%
      \relax%
    \else%
      \setlength{\unitlength}{\unitlength * \real{\svgscale}}%
    \fi%
  \else%
    \setlength{\unitlength}{\svgwidth}%
  \fi%
  \global\let\svgwidth\undefined%
  \global\let\svgscale\undefined%
  \makeatother%
  \begin{picture}(1,0.44235964)%
    \lineheight{1}%
    \setlength\tabcolsep{0pt}%
    \put(0,0){\includegraphics[width=\unitlength,page=1]{time_improvement_confident1.pdf}}%
    \put(0.03367651,0.02289768){\color[rgb]{0,0,0}\rotatebox{45.00000162}{\makebox(0,0)[lt]{\lineheight{1.25}\smash{\begin{tabular}[t]{l}motorc. lane change 1\end{tabular}}}}}%
    \put(0,0){\includegraphics[width=\unitlength,page=2]{time_improvement_confident1.pdf}}%
    \put(0.07332085,0.02289768){\color[rgb]{0,0,0}\rotatebox{45.00000162}{\makebox(0,0)[lt]{\lineheight{1.25}\smash{\begin{tabular}[t]{l}motorc. lane change 2\end{tabular}}}}}%
    \put(0,0){\includegraphics[width=\unitlength,page=3]{time_improvement_confident1.pdf}}%
    \put(0.1129652,0.02289769){\color[rgb]{0,0,0}\rotatebox{45.00000162}{\makebox(0,0)[lt]{\lineheight{1.25}\smash{\begin{tabular}[t]{l}motorc. lane change 3\end{tabular}}}}}%
    \put(0,0){\includegraphics[width=\unitlength,page=4]{time_improvement_confident1.pdf}}%
    \put(0.19877112,0.06840386){\color[rgb]{0,0,0}\rotatebox{45.00000162}{\makebox(0,0)[lt]{\lineheight{1.25}\smash{\begin{tabular}[t]{l}car overtaking 1\end{tabular}}}}}%
    \put(0,0){\includegraphics[width=\unitlength,page=5]{time_improvement_confident1.pdf}}%
    \put(0.23841546,0.06840386){\color[rgb]{0,0,0}\rotatebox{45.00000162}{\makebox(0,0)[lt]{\lineheight{1.25}\smash{\begin{tabular}[t]{l}car overtaking 2\end{tabular}}}}}%
    \put(0,0){\includegraphics[width=\unitlength,page=6]{time_improvement_confident1.pdf}}%
    \put(0.2780598,0.06840386){\color[rgb]{0,0,0}\rotatebox{45.00000162}{\makebox(0,0)[lt]{\lineheight{1.25}\smash{\begin{tabular}[t]{l}car overtaking 3\end{tabular}}}}}%
    \put(0,0){\includegraphics[width=\unitlength,page=7]{time_improvement_confident1.pdf}}%
    \put(0.27673091,0.02677524){\color[rgb]{0,0,0}\rotatebox{45.00000162}{\makebox(0,0)[lt]{\lineheight{1.25}\smash{\begin{tabular}[t]{l}priority intersection 1\end{tabular}}}}}%
    \put(0,0){\includegraphics[width=\unitlength,page=8]{time_improvement_confident1.pdf}}%
    \put(0.31637525,0.02677523){\color[rgb]{0,0,0}\rotatebox{45.00000162}{\makebox(0,0)[lt]{\lineheight{1.25}\smash{\begin{tabular}[t]{l}priority intersection 2\end{tabular}}}}}%
    \put(0,0){\includegraphics[width=\unitlength,page=9]{time_improvement_confident1.pdf}}%
    \put(0.35601959,0.02677523){\color[rgb]{0,0,0}\rotatebox{45.00000162}{\makebox(0,0)[lt]{\lineheight{1.25}\smash{\begin{tabular}[t]{l}priority intersection 3\end{tabular}}}}}%
    \put(0,0){\includegraphics[width=\unitlength,page=10]{time_improvement_confident1.pdf}}%
    \put(0.35680702,-0.01208167){\color[rgb]{0,0,0}\rotatebox{45.00000162}{\makebox(0,0)[lt]{\lineheight{1.25}\smash{\begin{tabular}[t]{l}curve taking intersection 1\end{tabular}}}}}%
    \put(0,0){\includegraphics[width=\unitlength,page=11]{time_improvement_confident1.pdf}}%
    \put(0.39645137,-0.01208166){\color[rgb]{0,0,0}\rotatebox{45.00000162}{\makebox(0,0)[lt]{\lineheight{1.25}\smash{\begin{tabular}[t]{l}curve taking intersection 2\end{tabular}}}}}%
    \put(0,0){\includegraphics[width=\unitlength,page=12]{time_improvement_confident1.pdf}}%
    \put(0.43609571,-0.01208167){\color[rgb]{0,0,0}\rotatebox{45.00000162}{\makebox(0,0)[lt]{\lineheight{1.25}\smash{\begin{tabular}[t]{l}curve taking intersection 3\end{tabular}}}}}%
    \put(0,0){\includegraphics[width=\unitlength,page=13]{time_improvement_confident1.pdf}}%
    \put(0.56158347,0.07376176){\color[rgb]{0,0,0}\rotatebox{45.00000162}{\makebox(0,0)[lt]{\lineheight{1.25}\smash{\begin{tabular}[t]{l}bicycle cut-in 1\end{tabular}}}}}%
    \put(0,0){\includegraphics[width=\unitlength,page=14]{time_improvement_confident1.pdf}}%
    \put(0.60122781,0.07376176){\color[rgb]{0,0,0}\rotatebox{45.00000162}{\makebox(0,0)[lt]{\lineheight{1.25}\smash{\begin{tabular}[t]{l}bicycle cut-in 2\end{tabular}}}}}%
    \put(0,0){\includegraphics[width=\unitlength,page=15]{time_improvement_confident1.pdf}}%
    \put(0.64087215,0.07376176){\color[rgb]{0,0,0}\rotatebox{45.00000162}{\makebox(0,0)[lt]{\lineheight{1.25}\smash{\begin{tabular}[t]{l}bicycle cut-in 3\end{tabular}}}}}%
    \put(0,0){\includegraphics[width=\unitlength,page=16]{time_improvement_confident1.pdf}}%
    \put(0.65592446,0.05179129){\color[rgb]{0,0,0}\rotatebox{45.00000162}{\makebox(0,0)[lt]{\lineheight{1.25}\smash{\begin{tabular}[t]{l}pedestrian cut-in 1\end{tabular}}}}}%
    \put(0,0){\includegraphics[width=\unitlength,page=17]{time_improvement_confident1.pdf}}%
    \put(0.6955688,0.05179129){\color[rgb]{0,0,0}\rotatebox{45.00000162}{\makebox(0,0)[lt]{\lineheight{1.25}\smash{\begin{tabular}[t]{l}pedestrian cut-in 2\end{tabular}}}}}%
    \put(0,0){\includegraphics[width=\unitlength,page=18]{time_improvement_confident1.pdf}}%
    \put(0.73521315,0.0517913){\color[rgb]{0,0,0}\rotatebox{45.00000162}{\makebox(0,0)[lt]{\lineheight{1.25}\smash{\begin{tabular}[t]{l}pedestrian cut-in 3\end{tabular}}}}}%
    \put(0,0){\includegraphics[width=\unitlength,page=19]{time_improvement_confident1.pdf}}%
    \put(0.0016462,0.35383829){\color[rgb]{0,0,0}\makebox(0,0)[lt]{\lineheight{1.25}\smash{\begin{tabular}[t]{l}no driver error\end{tabular}}}}%
    \put(0,0){\includegraphics[width=\unitlength,page=20]{time_improvement_confident1.pdf}}%
    \put(0.03126569,0.31419397){\color[rgb]{0,0,0}\makebox(0,0)[lt]{\lineheight{1.25}\smash{\begin{tabular}[t]{l}notice error\end{tabular}}}}%
    \put(0,0){\includegraphics[width=\unitlength,page=21]{time_improvement_confident1.pdf}}%
    \put(0.0113245,0.27454962){\color[rgb]{0,0,0}\makebox(0,0)[lt]{\lineheight{1.25}\smash{\begin{tabular}[t]{l}forecast error\end{tabular}}}}%
    \put(0,0){\includegraphics[width=\unitlength,page=22]{time_improvement_confident1.pdf}}%
    \put(-0.0019588,0.23490527){\color[rgb]{0,0,0}\makebox(0,0)[lt]{\lineheight{1.25}\smash{\begin{tabular}[t]{l}inference error\end{tabular}}}}%
    \put(0,0){\includegraphics[width=\unitlength,page=23]{time_improvement_confident1.pdf}}%
    \put(0.95836384,0.23091084){\color[rgb]{0,0,0}\makebox(0,0)[lt]{\lineheight{1.25}\smash{\begin{tabular}[t]{l}4\end{tabular}}}}%
    \put(0,0){\includegraphics[width=\unitlength,page=24]{time_improvement_confident1.pdf}}%
    \put(0.95836384,0.26264131){\color[rgb]{0,0,0}\makebox(0,0)[lt]{\lineheight{1.25}\smash{\begin{tabular}[t]{l}2\end{tabular}}}}%
    \put(0,0){\includegraphics[width=\unitlength,page=25]{time_improvement_confident1.pdf}}%
    \put(0.94094777,0.29437179){\color[rgb]{0,0,0}\makebox(0,0)[lt]{\lineheight{1.25}\smash{\begin{tabular}[t]{l}0\end{tabular}}}}%
    \put(0,0){\includegraphics[width=\unitlength,page=26]{time_improvement_confident1.pdf}}%
    \put(0.94094777,0.32610227){\color[rgb]{0,0,0}\makebox(0,0)[lt]{\lineheight{1.25}\smash{\begin{tabular}[t]{l}2\end{tabular}}}}%
    \put(0,0){\includegraphics[width=\unitlength,page=27]{time_improvement_confident1.pdf}}%
    \put(0.94094777,0.35783275){\color[rgb]{0,0,0}\makebox(0,0)[lt]{\lineheight{1.25}\smash{\begin{tabular}[t]{l}4\end{tabular}}}}%
    \put(0.98420782,0.41205406){\color[rgb]{0,0,0}\rotatebox{-90}{\makebox(0,0)[lt]{\lineheight{1.25}\smash{\begin{tabular}[t]{l}warning time difference [s]\end{tabular}}}}}%
    \put(0,0){\includegraphics[width=\unitlength,page=28]{time_improvement_confident1.pdf}}%
    \put(0.44406496,0.39356184){\color[rgb]{0,0,0}\makebox(0,0)[lt]{\lineheight{1.25}\smash{\begin{tabular}[t]{l}confident driver\end{tabular}}}}%
  \end{picture}%
\endgroup%

%% file: images/experiments/error_improvement_defensive1.pdf_tex
\begingroup%
  \makeatletter%
  \providecommand\color[2][]{%
    \errmessage{(Inkscape) Color is used for the text in Inkscape, but the package 'color.sty' is not loaded}%
    \renewcommand\color[2][]{}%
  }%
  \providecommand\transparent[1]{%
    \errmessage{(Inkscape) Transparency is used (non-zero) for the text in Inkscape, but the package 'transparent.sty' is not loaded}%
    \renewcommand\transparent[1]{}%
  }%
  \providecommand\rotatebox[2]{#2}%
  \newcommand*\fsize{\dimexpr\f@size pt\relax}%
  \newcommand*\lineheight[1]{\fontsize{\fsize}{#1\fsize}\selectfont}%
  \ifx\svgwidth\undefined%
    \setlength{\unitlength}{370.10212191bp}%
    \ifx\svgscale\undefined%
      \relax%
    \else%
      \setlength{\unitlength}{\unitlength * \real{\svgscale}}%
    \fi%
  \else%
    \setlength{\unitlength}{\svgwidth}%
  \fi%
  \global\let\svgwidth\undefined%
  \global\let\svgscale\undefined%
  \makeatother%
  \begin{picture}(1,0.44235964)%
    \lineheight{1}%
    \setlength\tabcolsep{0pt}%
    \put(0,0){\includegraphics[width=\unitlength,page=1]{error_improvement_defensive1.pdf}}%
    \put(0.03367651,0.02289768){\color[rgb]{0,0,0}\rotatebox{45.00000162}{\makebox(0,0)[lt]{\lineheight{1.25}\smash{\begin{tabular}[t]{l}motorc. lane change 1\end{tabular}}}}}%
    \put(0,0){\includegraphics[width=\unitlength,page=2]{error_improvement_defensive1.pdf}}%
    \put(0.07332085,0.02289768){\color[rgb]{0,0,0}\rotatebox{45.00000162}{\makebox(0,0)[lt]{\lineheight{1.25}\smash{\begin{tabular}[t]{l}motorc. lane change 2\end{tabular}}}}}%
    \put(0,0){\includegraphics[width=\unitlength,page=3]{error_improvement_defensive1.pdf}}%
    \put(0.1129652,0.02289769){\color[rgb]{0,0,0}\rotatebox{45.00000162}{\makebox(0,0)[lt]{\lineheight{1.25}\smash{\begin{tabular}[t]{l}motorc. lane change 3\end{tabular}}}}}%
    \put(0,0){\includegraphics[width=\unitlength,page=4]{error_improvement_defensive1.pdf}}%
    \put(0.19877112,0.06840386){\color[rgb]{0,0,0}\rotatebox{45.00000162}{\makebox(0,0)[lt]{\lineheight{1.25}\smash{\begin{tabular}[t]{l}car overtaking 1\end{tabular}}}}}%
    \put(0,0){\includegraphics[width=\unitlength,page=5]{error_improvement_defensive1.pdf}}%
    \put(0.23841546,0.06840386){\color[rgb]{0,0,0}\rotatebox{45.00000162}{\makebox(0,0)[lt]{\lineheight{1.25}\smash{\begin{tabular}[t]{l}car overtaking 2\end{tabular}}}}}%
    \put(0,0){\includegraphics[width=\unitlength,page=6]{error_improvement_defensive1.pdf}}%
    \put(0.2780598,0.06840386){\color[rgb]{0,0,0}\rotatebox{45.00000162}{\makebox(0,0)[lt]{\lineheight{1.25}\smash{\begin{tabular}[t]{l}car overtaking 3\end{tabular}}}}}%
    \put(0,0){\includegraphics[width=\unitlength,page=7]{error_improvement_defensive1.pdf}}%
    \put(0.27673091,0.02677524){\color[rgb]{0,0,0}\rotatebox{45.00000162}{\makebox(0,0)[lt]{\lineheight{1.25}\smash{\begin{tabular}[t]{l}priority intersection 1\end{tabular}}}}}%
    \put(0,0){\includegraphics[width=\unitlength,page=8]{error_improvement_defensive1.pdf}}%
    \put(0.31637525,0.02677523){\color[rgb]{0,0,0}\rotatebox{45.00000162}{\makebox(0,0)[lt]{\lineheight{1.25}\smash{\begin{tabular}[t]{l}priority intersection 2\end{tabular}}}}}%
    \put(0,0){\includegraphics[width=\unitlength,page=9]{error_improvement_defensive1.pdf}}%
    \put(0.35601959,0.02677523){\color[rgb]{0,0,0}\rotatebox{45.00000162}{\makebox(0,0)[lt]{\lineheight{1.25}\smash{\begin{tabular}[t]{l}priority intersection 3\end{tabular}}}}}%
    \put(0,0){\includegraphics[width=\unitlength,page=10]{error_improvement_defensive1.pdf}}%
    \put(0.35680702,-0.01208167){\color[rgb]{0,0,0}\rotatebox{45.00000162}{\makebox(0,0)[lt]{\lineheight{1.25}\smash{\begin{tabular}[t]{l}curve taking intersection 1\end{tabular}}}}}%
    \put(0,0){\includegraphics[width=\unitlength,page=11]{error_improvement_defensive1.pdf}}%
    \put(0.39645137,-0.01208166){\color[rgb]{0,0,0}\rotatebox{45.00000162}{\makebox(0,0)[lt]{\lineheight{1.25}\smash{\begin{tabular}[t]{l}curve taking intersection 2\end{tabular}}}}}%
    \put(0,0){\includegraphics[width=\unitlength,page=12]{error_improvement_defensive1.pdf}}%
    \put(0.43609571,-0.01208167){\color[rgb]{0,0,0}\rotatebox{45.00000162}{\makebox(0,0)[lt]{\lineheight{1.25}\smash{\begin{tabular}[t]{l}curve taking intersection 3\end{tabular}}}}}%
    \put(0,0){\includegraphics[width=\unitlength,page=13]{error_improvement_defensive1.pdf}}%
    \put(0.56158347,0.07376176){\color[rgb]{0,0,0}\rotatebox{45.00000162}{\makebox(0,0)[lt]{\lineheight{1.25}\smash{\begin{tabular}[t]{l}bicycle cut-in 1\end{tabular}}}}}%
    \put(0,0){\includegraphics[width=\unitlength,page=14]{error_improvement_defensive1.pdf}}%
    \put(0.60122781,0.07376176){\color[rgb]{0,0,0}\rotatebox{45.00000162}{\makebox(0,0)[lt]{\lineheight{1.25}\smash{\begin{tabular}[t]{l}bicycle cut-in 2\end{tabular}}}}}%
    \put(0,0){\includegraphics[width=\unitlength,page=15]{error_improvement_defensive1.pdf}}%
    \put(0.64087215,0.07376176){\color[rgb]{0,0,0}\rotatebox{45.00000162}{\makebox(0,0)[lt]{\lineheight{1.25}\smash{\begin{tabular}[t]{l}bicycle cut-in 3\end{tabular}}}}}%
    \put(0,0){\includegraphics[width=\unitlength,page=16]{error_improvement_defensive1.pdf}}%
    \put(0.65592446,0.05179129){\color[rgb]{0,0,0}\rotatebox{45.00000162}{\makebox(0,0)[lt]{\lineheight{1.25}\smash{\begin{tabular}[t]{l}pedestrian cut-in 1\end{tabular}}}}}%
    \put(0,0){\includegraphics[width=\unitlength,page=17]{error_improvement_defensive1.pdf}}%
    \put(0.6955688,0.05179129){\color[rgb]{0,0,0}\rotatebox{45.00000162}{\makebox(0,0)[lt]{\lineheight{1.25}\smash{\begin{tabular}[t]{l}pedestrian cut-in 2\end{tabular}}}}}%
    \put(0,0){\includegraphics[width=\unitlength,page=18]{error_improvement_defensive1.pdf}}%
    \put(0.73521315,0.0517913){\color[rgb]{0,0,0}\rotatebox{45.00000162}{\makebox(0,0)[lt]{\lineheight{1.25}\smash{\begin{tabular}[t]{l}pedestrian cut-in 3\end{tabular}}}}}%
    \put(0,0){\includegraphics[width=\unitlength,page=19]{error_improvement_defensive1.pdf}}%
    \put(0.0016462,0.35383829){\color[rgb]{0,0,0}\makebox(0,0)[lt]{\lineheight{1.25}\smash{\begin{tabular}[t]{l}no driver error\end{tabular}}}}%
    \put(0,0){\includegraphics[width=\unitlength,page=20]{error_improvement_defensive1.pdf}}%
    \put(0.03126569,0.31419397){\color[rgb]{0,0,0}\makebox(0,0)[lt]{\lineheight{1.25}\smash{\begin{tabular}[t]{l}notice error\end{tabular}}}}%
    \put(0,0){\includegraphics[width=\unitlength,page=21]{error_improvement_defensive1.pdf}}%
    \put(0.0113245,0.27454962){\color[rgb]{0,0,0}\makebox(0,0)[lt]{\lineheight{1.25}\smash{\begin{tabular}[t]{l}forecast error\end{tabular}}}}%
    \put(0,0){\includegraphics[width=\unitlength,page=22]{error_improvement_defensive1.pdf}}%
    \put(-0.0019588,0.23490527){\color[rgb]{0,0,0}\makebox(0,0)[lt]{\lineheight{1.25}\smash{\begin{tabular}[t]{l}inference error\end{tabular}}}}%
    \put(0,0){\includegraphics[width=\unitlength,page=23]{error_improvement_defensive1.pdf}}%
    \put(0.44189669,0.39352322){\color[rgb]{0,0,0}\makebox(0,0)[lt]{\lineheight{1.25}\smash{\begin{tabular}[t]{l}defensive driver\end{tabular}}}}%
    \put(0,0){\includegraphics[width=\unitlength,page=24]{error_improvement_defensive1.pdf}}%
    \put(0.94105202,0.2144774){\color[rgb]{0,0,0}\makebox(0,0)[lt]{\lineheight{1.25}\smash{\begin{tabular}[t]{l}0\end{tabular}}}}%
    \put(0,0){\includegraphics[width=\unitlength,page=25]{error_improvement_defensive1.pdf}}%
    \put(0.94105202,0.29560113){\color[rgb]{0,0,0}\makebox(0,0)[lt]{\lineheight{1.25}\smash{\begin{tabular}[t]{l}1\end{tabular}}}}%
    \put(0,0){\includegraphics[width=\unitlength,page=26]{error_improvement_defensive1.pdf}}%
    \put(0.94105202,0.37672486){\color[rgb]{0,0,0}\makebox(0,0)[lt]{\lineheight{1.25}\smash{\begin{tabular}[t]{l}2\end{tabular}}}}%
    \put(0.9669041,0.41060398){\color[rgb]{0,0,0}\rotatebox{-90}{\makebox(0,0)[lt]{\lineheight{1.25}\smash{\begin{tabular}[t]{l}false positive (1), false negative (2)\end{tabular}}}}}%
    \put(0,0){\includegraphics[width=\unitlength,page=27]{error_improvement_defensive1.pdf}}%
  \end{picture}%
\endgroup%

%% file: images/experiments/error_improvement_normal1.pdf_tex
\begingroup%
  \makeatletter%
  \providecommand\color[2][]{%
    \errmessage{(Inkscape) Color is used for the text in Inkscape, but the package 'color.sty' is not loaded}%
    \renewcommand\color[2][]{}%
  }%
  \providecommand\transparent[1]{%
    \errmessage{(Inkscape) Transparency is used (non-zero) for the text in Inkscape, but the package 'transparent.sty' is not loaded}%
    \renewcommand\transparent[1]{}%
  }%
  \providecommand\rotatebox[2]{#2}%
  \newcommand*\fsize{\dimexpr\f@size pt\relax}%
  \newcommand*\lineheight[1]{\fontsize{\fsize}{#1\fsize}\selectfont}%
  \ifx\svgwidth\undefined%
    \setlength{\unitlength}{370.10212191bp}%
    \ifx\svgscale\undefined%
      \relax%
    \else%
      \setlength{\unitlength}{\unitlength * \real{\svgscale}}%
    \fi%
  \else%
    \setlength{\unitlength}{\svgwidth}%
  \fi%
  \global\let\svgwidth\undefined%
  \global\let\svgscale\undefined%
  \makeatother%
  \begin{picture}(1,0.44235964)%
    \lineheight{1}%
    \setlength\tabcolsep{0pt}%
    \put(0,0){\includegraphics[width=\unitlength,page=1]{error_improvement_normal1.pdf}}%
    \put(0.03367651,0.02289768){\color[rgb]{0,0,0}\rotatebox{45.00000162}{\makebox(0,0)[lt]{\lineheight{1.25}\smash{\begin{tabular}[t]{l}motorc. lane change 1\end{tabular}}}}}%
    \put(0,0){\includegraphics[width=\unitlength,page=2]{error_improvement_normal1.pdf}}%
    \put(0.07332085,0.02289768){\color[rgb]{0,0,0}\rotatebox{45.00000162}{\makebox(0,0)[lt]{\lineheight{1.25}\smash{\begin{tabular}[t]{l}motorc. lane change 2\end{tabular}}}}}%
    \put(0,0){\includegraphics[width=\unitlength,page=3]{error_improvement_normal1.pdf}}%
    \put(0.1129652,0.02289769){\color[rgb]{0,0,0}\rotatebox{45.00000162}{\makebox(0,0)[lt]{\lineheight{1.25}\smash{\begin{tabular}[t]{l}motorc. lane change 3\end{tabular}}}}}%
    \put(0,0){\includegraphics[width=\unitlength,page=4]{error_improvement_normal1.pdf}}%
    \put(0.19877112,0.06840386){\color[rgb]{0,0,0}\rotatebox{45.00000162}{\makebox(0,0)[lt]{\lineheight{1.25}\smash{\begin{tabular}[t]{l}car overtaking 1\end{tabular}}}}}%
    \put(0,0){\includegraphics[width=\unitlength,page=5]{error_improvement_normal1.pdf}}%
    \put(0.23841546,0.06840386){\color[rgb]{0,0,0}\rotatebox{45.00000162}{\makebox(0,0)[lt]{\lineheight{1.25}\smash{\begin{tabular}[t]{l}car overtaking 2\end{tabular}}}}}%
    \put(0,0){\includegraphics[width=\unitlength,page=6]{error_improvement_normal1.pdf}}%
    \put(0.2780598,0.06840386){\color[rgb]{0,0,0}\rotatebox{45.00000162}{\makebox(0,0)[lt]{\lineheight{1.25}\smash{\begin{tabular}[t]{l}car overtaking 3\end{tabular}}}}}%
    \put(0,0){\includegraphics[width=\unitlength,page=7]{error_improvement_normal1.pdf}}%
    \put(0.27673091,0.02677524){\color[rgb]{0,0,0}\rotatebox{45.00000162}{\makebox(0,0)[lt]{\lineheight{1.25}\smash{\begin{tabular}[t]{l}priority intersection 1\end{tabular}}}}}%
    \put(0,0){\includegraphics[width=\unitlength,page=8]{error_improvement_normal1.pdf}}%
    \put(0.31637525,0.02677523){\color[rgb]{0,0,0}\rotatebox{45.00000162}{\makebox(0,0)[lt]{\lineheight{1.25}\smash{\begin{tabular}[t]{l}priority intersection 2\end{tabular}}}}}%
    \put(0,0){\includegraphics[width=\unitlength,page=9]{error_improvement_normal1.pdf}}%
    \put(0.35601959,0.02677523){\color[rgb]{0,0,0}\rotatebox{45.00000162}{\makebox(0,0)[lt]{\lineheight{1.25}\smash{\begin{tabular}[t]{l}priority intersection 3\end{tabular}}}}}%
    \put(0,0){\includegraphics[width=\unitlength,page=10]{error_improvement_normal1.pdf}}%
    \put(0.35680702,-0.01208167){\color[rgb]{0,0,0}\rotatebox{45.00000162}{\makebox(0,0)[lt]{\lineheight{1.25}\smash{\begin{tabular}[t]{l}curve taking intersection 1\end{tabular}}}}}%
    \put(0,0){\includegraphics[width=\unitlength,page=11]{error_improvement_normal1.pdf}}%
    \put(0.39645137,-0.01208166){\color[rgb]{0,0,0}\rotatebox{45.00000162}{\makebox(0,0)[lt]{\lineheight{1.25}\smash{\begin{tabular}[t]{l}curve taking intersection 2\end{tabular}}}}}%
    \put(0,0){\includegraphics[width=\unitlength,page=12]{error_improvement_normal1.pdf}}%
    \put(0.43609571,-0.01208167){\color[rgb]{0,0,0}\rotatebox{45.00000162}{\makebox(0,0)[lt]{\lineheight{1.25}\smash{\begin{tabular}[t]{l}curve taking intersection 3\end{tabular}}}}}%
    \put(0,0){\includegraphics[width=\unitlength,page=13]{error_improvement_normal1.pdf}}%
    \put(0.56158347,0.07376176){\color[rgb]{0,0,0}\rotatebox{45.00000162}{\makebox(0,0)[lt]{\lineheight{1.25}\smash{\begin{tabular}[t]{l}bicycle cut-in 1\end{tabular}}}}}%
    \put(0,0){\includegraphics[width=\unitlength,page=14]{error_improvement_normal1.pdf}}%
    \put(0.60122781,0.07376176){\color[rgb]{0,0,0}\rotatebox{45.00000162}{\makebox(0,0)[lt]{\lineheight{1.25}\smash{\begin{tabular}[t]{l}bicycle cut-in 2\end{tabular}}}}}%
    \put(0,0){\includegraphics[width=\unitlength,page=15]{error_improvement_normal1.pdf}}%
    \put(0.64087215,0.07376176){\color[rgb]{0,0,0}\rotatebox{45.00000162}{\makebox(0,0)[lt]{\lineheight{1.25}\smash{\begin{tabular}[t]{l}bicycle cut-in 3\end{tabular}}}}}%
    \put(0,0){\includegraphics[width=\unitlength,page=16]{error_improvement_normal1.pdf}}%
    \put(0.65592446,0.05179129){\color[rgb]{0,0,0}\rotatebox{45.00000162}{\makebox(0,0)[lt]{\lineheight{1.25}\smash{\begin{tabular}[t]{l}pedestrian cut-in 1\end{tabular}}}}}%
    \put(0,0){\includegraphics[width=\unitlength,page=17]{error_improvement_normal1.pdf}}%
    \put(0.6955688,0.05179129){\color[rgb]{0,0,0}\rotatebox{45.00000162}{\makebox(0,0)[lt]{\lineheight{1.25}\smash{\begin{tabular}[t]{l}pedestrian cut-in 2\end{tabular}}}}}%
    \put(0,0){\includegraphics[width=\unitlength,page=18]{error_improvement_normal1.pdf}}%
    \put(0.73521315,0.0517913){\color[rgb]{0,0,0}\rotatebox{45.00000162}{\makebox(0,0)[lt]{\lineheight{1.25}\smash{\begin{tabular}[t]{l}pedestrian cut-in 3\end{tabular}}}}}%
    \put(0,0){\includegraphics[width=\unitlength,page=19]{error_improvement_normal1.pdf}}%
    \put(0.0016462,0.35383829){\color[rgb]{0,0,0}\makebox(0,0)[lt]{\lineheight{1.25}\smash{\begin{tabular}[t]{l}no driver error\end{tabular}}}}%
    \put(0,0){\includegraphics[width=\unitlength,page=20]{error_improvement_normal1.pdf}}%
    \put(0.03126569,0.31419397){\color[rgb]{0,0,0}\makebox(0,0)[lt]{\lineheight{1.25}\smash{\begin{tabular}[t]{l}notice error\end{tabular}}}}%
    \put(0,0){\includegraphics[width=\unitlength,page=21]{error_improvement_normal1.pdf}}%
    \put(0.0113245,0.27454962){\color[rgb]{0,0,0}\makebox(0,0)[lt]{\lineheight{1.25}\smash{\begin{tabular}[t]{l}forecast error\end{tabular}}}}%
    \put(0,0){\includegraphics[width=\unitlength,page=22]{error_improvement_normal1.pdf}}%
    \put(-0.0019588,0.23490527){\color[rgb]{0,0,0}\makebox(0,0)[lt]{\lineheight{1.25}\smash{\begin{tabular}[t]{l}inference error\end{tabular}}}}%
    \put(0,0){\includegraphics[width=\unitlength,page=23]{error_improvement_normal1.pdf}}%
    \put(0.94105202,0.2144774){\color[rgb]{0,0,0}\makebox(0,0)[lt]{\lineheight{1.25}\smash{\begin{tabular}[t]{l}0\end{tabular}}}}%
    \put(0,0){\includegraphics[width=\unitlength,page=24]{error_improvement_normal1.pdf}}%
    \put(0.94105202,0.29560113){\color[rgb]{0,0,0}\makebox(0,0)[lt]{\lineheight{1.25}\smash{\begin{tabular}[t]{l}1\end{tabular}}}}%
    \put(0,0){\includegraphics[width=\unitlength,page=25]{error_improvement_normal1.pdf}}%
    \put(0.94105202,0.37672486){\color[rgb]{0,0,0}\makebox(0,0)[lt]{\lineheight{1.25}\smash{\begin{tabular}[t]{l}2\end{tabular}}}}%
    \put(0.9669041,0.41060398){\color[rgb]{0,0,0}\rotatebox{-90}{\makebox(0,0)[lt]{\lineheight{1.25}\smash{\begin{tabular}[t]{l}false positive (1), false negative (2)\end{tabular}}}}}%
    \put(0,0){\includegraphics[width=\unitlength,page=26]{error_improvement_normal1.pdf}}%
    \put(0.46047787,0.39365498){\color[rgb]{0,0,0}\makebox(0,0)[lt]{\lineheight{1.25}\smash{\begin{tabular}[t]{l}normal driver\end{tabular}}}}%
  \end{picture}%
\endgroup%

%% file: images/experiments/error_improvement_confident1.pdf_tex
\begingroup%
  \makeatletter%
  \providecommand\color[2][]{%
    \errmessage{(Inkscape) Color is used for the text in Inkscape, but the package 'color.sty' is not loaded}%
    \renewcommand\color[2][]{}%
  }%
  \providecommand\transparent[1]{%
    \errmessage{(Inkscape) Transparency is used (non-zero) for the text in Inkscape, but the package 'transparent.sty' is not loaded}%
    \renewcommand\transparent[1]{}%
  }%
  \providecommand\rotatebox[2]{#2}%
  \newcommand*\fsize{\dimexpr\f@size pt\relax}%
  \newcommand*\lineheight[1]{\fontsize{\fsize}{#1\fsize}\selectfont}%
  \ifx\svgwidth\undefined%
    \setlength{\unitlength}{370.10212191bp}%
    \ifx\svgscale\undefined%
      \relax%
    \else%
      \setlength{\unitlength}{\unitlength * \real{\svgscale}}%
    \fi%
  \else%
    \setlength{\unitlength}{\svgwidth}%
  \fi%
  \global\let\svgwidth\undefined%
  \global\let\svgscale\undefined%
  \makeatother%
  \begin{picture}(1,0.44235964)%
    \lineheight{1}%
    \setlength\tabcolsep{0pt}%
    \put(0,0){\includegraphics[width=\unitlength,page=1]{error_improvement_confident1.pdf}}%
    \put(0.03367651,0.02289768){\color[rgb]{0,0,0}\rotatebox{45.00000162}{\makebox(0,0)[lt]{\lineheight{1.25}\smash{\begin{tabular}[t]{l}motorc. lane change 1\end{tabular}}}}}%
    \put(0,0){\includegraphics[width=\unitlength,page=2]{error_improvement_confident1.pdf}}%
    \put(0.07332085,0.02289768){\color[rgb]{0,0,0}\rotatebox{45.00000162}{\makebox(0,0)[lt]{\lineheight{1.25}\smash{\begin{tabular}[t]{l}motorc. lane change 2\end{tabular}}}}}%
    \put(0,0){\includegraphics[width=\unitlength,page=3]{error_improvement_confident1.pdf}}%
    \put(0.1129652,0.02289769){\color[rgb]{0,0,0}\rotatebox{45.00000162}{\makebox(0,0)[lt]{\lineheight{1.25}\smash{\begin{tabular}[t]{l}motorc. lane change 3\end{tabular}}}}}%
    \put(0,0){\includegraphics[width=\unitlength,page=4]{error_improvement_confident1.pdf}}%
    \put(0.19877112,0.06840386){\color[rgb]{0,0,0}\rotatebox{45.00000162}{\makebox(0,0)[lt]{\lineheight{1.25}\smash{\begin{tabular}[t]{l}car overtaking 1\end{tabular}}}}}%
    \put(0,0){\includegraphics[width=\unitlength,page=5]{error_improvement_confident1.pdf}}%
    \put(0.23841546,0.06840386){\color[rgb]{0,0,0}\rotatebox{45.00000162}{\makebox(0,0)[lt]{\lineheight{1.25}\smash{\begin{tabular}[t]{l}car overtaking 2\end{tabular}}}}}%
    \put(0,0){\includegraphics[width=\unitlength,page=6]{error_improvement_confident1.pdf}}%
    \put(0.2780598,0.06840386){\color[rgb]{0,0,0}\rotatebox{45.00000162}{\makebox(0,0)[lt]{\lineheight{1.25}\smash{\begin{tabular}[t]{l}car overtaking 3\end{tabular}}}}}%
    \put(0,0){\includegraphics[width=\unitlength,page=7]{error_improvement_confident1.pdf}}%
    \put(0.27673091,0.02677524){\color[rgb]{0,0,0}\rotatebox{45.00000162}{\makebox(0,0)[lt]{\lineheight{1.25}\smash{\begin{tabular}[t]{l}priority intersection 1\end{tabular}}}}}%
    \put(0,0){\includegraphics[width=\unitlength,page=8]{error_improvement_confident1.pdf}}%
    \put(0.31637525,0.02677523){\color[rgb]{0,0,0}\rotatebox{45.00000162}{\makebox(0,0)[lt]{\lineheight{1.25}\smash{\begin{tabular}[t]{l}priority intersection 2\end{tabular}}}}}%
    \put(0,0){\includegraphics[width=\unitlength,page=9]{error_improvement_confident1.pdf}}%
    \put(0.35601959,0.02677523){\color[rgb]{0,0,0}\rotatebox{45.00000162}{\makebox(0,0)[lt]{\lineheight{1.25}\smash{\begin{tabular}[t]{l}priority intersection 3\end{tabular}}}}}%
    \put(0,0){\includegraphics[width=\unitlength,page=10]{error_improvement_confident1.pdf}}%
    \put(0.35680702,-0.01208167){\color[rgb]{0,0,0}\rotatebox{45.00000162}{\makebox(0,0)[lt]{\lineheight{1.25}\smash{\begin{tabular}[t]{l}curve taking intersection 1\end{tabular}}}}}%
    \put(0,0){\includegraphics[width=\unitlength,page=11]{error_improvement_confident1.pdf}}%
    \put(0.39645137,-0.01208166){\color[rgb]{0,0,0}\rotatebox{45.00000162}{\makebox(0,0)[lt]{\lineheight{1.25}\smash{\begin{tabular}[t]{l}curve taking intersection 2\end{tabular}}}}}%
    \put(0,0){\includegraphics[width=\unitlength,page=12]{error_improvement_confident1.pdf}}%
    \put(0.43609571,-0.01208167){\color[rgb]{0,0,0}\rotatebox{45.00000162}{\makebox(0,0)[lt]{\lineheight{1.25}\smash{\begin{tabular}[t]{l}curve taking intersection 3\end{tabular}}}}}%
    \put(0,0){\includegraphics[width=\unitlength,page=13]{error_improvement_confident1.pdf}}%
    \put(0.56158347,0.07376176){\color[rgb]{0,0,0}\rotatebox{45.00000162}{\makebox(0,0)[lt]{\lineheight{1.25}\smash{\begin{tabular}[t]{l}bicycle cut-in 1\end{tabular}}}}}%
    \put(0,0){\includegraphics[width=\unitlength,page=14]{error_improvement_confident1.pdf}}%
    \put(0.60122781,0.07376176){\color[rgb]{0,0,0}\rotatebox{45.00000162}{\makebox(0,0)[lt]{\lineheight{1.25}\smash{\begin{tabular}[t]{l}bicycle cut-in 2\end{tabular}}}}}%
    \put(0,0){\includegraphics[width=\unitlength,page=15]{error_improvement_confident1.pdf}}%
    \put(0.64087215,0.07376176){\color[rgb]{0,0,0}\rotatebox{45.00000162}{\makebox(0,0)[lt]{\lineheight{1.25}\smash{\begin{tabular}[t]{l}bicycle cut-in 3\end{tabular}}}}}%
    \put(0,0){\includegraphics[width=\unitlength,page=16]{error_improvement_confident1.pdf}}%
    \put(0.65592446,0.05179129){\color[rgb]{0,0,0}\rotatebox{45.00000162}{\makebox(0,0)[lt]{\lineheight{1.25}\smash{\begin{tabular}[t]{l}pedestrian cut-in 1\end{tabular}}}}}%
    \put(0,0){\includegraphics[width=\unitlength,page=17]{error_improvement_confident1.pdf}}%
    \put(0.6955688,0.05179129){\color[rgb]{0,0,0}\rotatebox{45.00000162}{\makebox(0,0)[lt]{\lineheight{1.25}\smash{\begin{tabular}[t]{l}pedestrian cut-in 2\end{tabular}}}}}%
    \put(0,0){\includegraphics[width=\unitlength,page=18]{error_improvement_confident1.pdf}}%
    \put(0.73521315,0.0517913){\color[rgb]{0,0,0}\rotatebox{45.00000162}{\makebox(0,0)[lt]{\lineheight{1.25}\smash{\begin{tabular}[t]{l}pedestrian cut-in 3\end{tabular}}}}}%
    \put(0,0){\includegraphics[width=\unitlength,page=19]{error_improvement_confident1.pdf}}%
    \put(0.0016462,0.35383829){\color[rgb]{0,0,0}\makebox(0,0)[lt]{\lineheight{1.25}\smash{\begin{tabular}[t]{l}no driver error\end{tabular}}}}%
    \put(0,0){\includegraphics[width=\unitlength,page=20]{error_improvement_confident1.pdf}}%
    \put(0.03126569,0.31419397){\color[rgb]{0,0,0}\makebox(0,0)[lt]{\lineheight{1.25}\smash{\begin{tabular}[t]{l}notice error\end{tabular}}}}%
    \put(0,0){\includegraphics[width=\unitlength,page=21]{error_improvement_confident1.pdf}}%
    \put(0.0113245,0.27454962){\color[rgb]{0,0,0}\makebox(0,0)[lt]{\lineheight{1.25}\smash{\begin{tabular}[t]{l}forecast error\end{tabular}}}}%
    \put(0,0){\includegraphics[width=\unitlength,page=22]{error_improvement_confident1.pdf}}%
    \put(-0.0019588,0.23490527){\color[rgb]{0,0,0}\makebox(0,0)[lt]{\lineheight{1.25}\smash{\begin{tabular}[t]{l}inference error\end{tabular}}}}%
    \put(0,0){\includegraphics[width=\unitlength,page=23]{error_improvement_confident1.pdf}}%
    \put(0.94105202,0.2144774){\color[rgb]{0,0,0}\makebox(0,0)[lt]{\lineheight{1.25}\smash{\begin{tabular}[t]{l}0\end{tabular}}}}%
    \put(0,0){\includegraphics[width=\unitlength,page=24]{error_improvement_confident1.pdf}}%
    \put(0.94105202,0.29560113){\color[rgb]{0,0,0}\makebox(0,0)[lt]{\lineheight{1.25}\smash{\begin{tabular}[t]{l}1\end{tabular}}}}%
    \put(0,0){\includegraphics[width=\unitlength,page=25]{error_improvement_confident1.pdf}}%
    \put(0.94105202,0.37672486){\color[rgb]{0,0,0}\makebox(0,0)[lt]{\lineheight{1.25}\smash{\begin{tabular}[t]{l}2\end{tabular}}}}%
    \put(0.9669041,0.41060398){\color[rgb]{0,0,0}\rotatebox{-90}{\makebox(0,0)[lt]{\lineheight{1.25}\smash{\begin{tabular}[t]{l}false positive (1), false negative (2)\end{tabular}}}}}%
    \put(0,0){\includegraphics[width=\unitlength,page=26]{error_improvement_confident1.pdf}}%
    \put(0.44426749,0.39357021){\color[rgb]{0,0,0}\makebox(0,0)[lt]{\lineheight{1.25}\smash{\begin{tabular}[t]{l}confident driver\end{tabular}}}}%
  \end{picture}%
\endgroup%

%% file: chapters/conclusion.tex
\section{Conclusion and Outlook}
\label{sec:conclusion}

In summary, in this paper, we presented a human-based risk model that considers the human driver in driver support. The proposed risk model combines a) the current driver perception with a so-called perceived Risk Maps planner and b) driver personalization with a personalized risk factor. Perceived Risk Maps considers here human-specific driver errors in the behavior planning and the personalization component additionally adapts the warning signal to the given driver type by changing the parametrization. 
 
In experiments, the human-based risk model was compared to the baseline risk model from \cite{puphal2019} that does not consider the human driver. In simulation variations of six different interactive driving scenarios, the proposed approach was shown to achieve improved driver support. On average, an earlier warning time of 1.55 seconds was achieved for a defensive driver and a warning error reduction of 28 $\%$ was achieved for a confident driver. The results highlight the generalizability of the approach to different interactive driving scenarios. In comparison to related work, the proposed human-based risk model is the first model that combines driver perception with driver personalization for improved driver support. 

A limitation of the work is that the experiments were done in simulation. In future work, we therefore intend to test the \makebox[\linewidth][s]{human-based risk model of this paper on real test vehicles}

\newpage 

\noindent in in order to investigate noise effects and error effects from sensed human driver information on the proposed risk model. Moreover, we aim to analyze if the driver prefers the human-based risk model in real vehicle drives. A user study helps to further fine-tune the parameters of the approach. Closing the simulation-to-reality gap of human factors is an essential research direction.